\documentclass[a4paper,11pt]{article}
\usepackage{a4wide}
\usepackage{graphicx}
\usepackage{subcaption}
\usepackage{xcolor}
\usepackage{amsmath,amsfonts,amssymb,amstext,graphicx}
\usepackage{placeins}
\usepackage[colorlinks=true,  citecolor=blue, linkcolor=blue, urlcolor=black]{hyperref}
\usepackage[numbers,sort&compress]{natbib}
\usepackage{breakurl}

\usepackage{cancel}
\usepackage{empheq}
\usepackage{float} 
\usepackage{subcaption}
\usepackage{authblk}
\usepackage{comment}
\usepackage{blindtext}
\usepackage{cases}
\usepackage{ulem}

\usepackage[bottom]{footmisc}
\usepackage{lineno}

\usepackage{tikz}
\usetikzlibrary{arrows, positioning}

\newcommand{\dv}[2]{\frac{d#1}{d#2}}

\def\bra#1{\mathinner{\langle{#1}|}}
\def\ket#1{\mathinner{|{#1}\rangle}}

\newcommand{\com}[2]{\big{[}#1,#2\big{]} }
\newcommand{\bigcom}[2]{\Big{[}#1,#2\Big{]} }

\newcommand{\trd}[2]{{{\rm Tr}_{#1} {#2}}}

\newcommand{\av}[1]{\langle #1 \rangle}

\begin{document} 
 \date{}
\title{\centerline \textbf {
%Cross-Correlated Environmental Fluctuations in Multichannel System–Bath Interactions
Effect of Cross-Spectral Correlations on Qubit Dynamics: Coherence Revival and Relaxation Modulation}}
\bigskip
\author{Siddhartha Dutta \thanks{siddhartha.dutta.phy23@gm.rkmvu.ac.in}}
\author{Sujay Mondal \thanks{sujay.mondal.phy23@gm.rkmvu.ac.in}}
\author{Abhijit Bandyopadhyay \thanks{abhijit.phy@gm.rkmvu.ac.in}}
\normalsize

\affil{Department of Physics\\
Ramakrishna Mission Vivekananda Educational and Research Institute\\ Belur Math, Howrah-711202, West-Bengal, India}
\date{\today}
\maketitle

\maketitle
 
\begin{abstract}
We investigate the reduced dynamics of a qubit subject to correlated longitudinal and transverse noise arising from its coupling to a shared bosonic bath.  The environmental fluctuations are characterized by a positive-semidefinite matrix-valued spectral density, whose complex off-diagonal elements encode correlations between dephasing and relaxation channels in the frequency domain.
 Within the second-order time-convolutionless framework, we derive closed time-local equations for the Bloch-vector components of the reduced density matrix. The numerical implementation is validated against the exact pure-dephasing solution and the established  behavior of the transverse-coupling spin-boson model. When both noise channels are present, the cross-spectral terms couple the otherwise distinct dephasing and relaxation sectors, producing dynamics that cannot be reproduced by adding independent noise contributions. In particular, the correlations generate non-monotonic population relaxation and a transient revival of coherence following its initial decay. The strength, bandwidth, delay, and phase of the cross spectrum provide control parameters for the magnitude and temporal structure of these effects. Our results demonstrate that correlated multi-axis noise can redistribute coherence loss and energy relaxation in time, thereby providing finite temporal windows of enhanced coherence or suppressed relaxation within the weak-coupling regime.
\end{abstract}

%\tableofcontents

\section{Introduction}
\label{sec:intro}
In the ideal limit of perfect isolation, state
of a quantum state  evolves coherently under unitary dynamics. For realistic quantum systems, this
coherent evolution  is modified by the system's
unavoidable coupling to environmental degrees of freedom
interaction with environmental degrees of freedom. 
Such interactions generally produce decoherence through the loss of phase information and dissipation through the exchange of energy between the system and its surroundings. Understanding these processes is essential for the characterization and control of quantum devices, where environmental noise limits the preservation, manipulation, and readout of quantum states \cite{Breuer2002,Weiss2012,RivasHuelga2012}.\\

For a two-level system, environmental fluctuations are commonly separated into longitudinal and transverse components. Longitudinal noise modulates the energy-level splitting and primarily induces pure dephasing, suppressing coherence in the energy eigenbasis without directly changing the populations. Transverse noise, on the other hand, couples different energy eigenstates and produces excitation and relaxation through energy exchange with the environment.
The  mechanisms of
longitudinal dephasing and transverse 
relaxation are commonly treated as 
statistically independent noise channels,
with their effects characterized separately by the corresponding dephasing and relaxation rates \cite{Makhlin2001,Ithier2005,Paladino2014,Clerk2010}.\\

In many realistic scenarios, environmental noise 
cannot, in general, be represented by statistically 
independent pure-dephasing and relaxation channels. A single 
microscopic noise source may couple to multiple non-commuting 
qubit operators, producing both fluctuations of the energy 
splitting and transitions between energy eigenstates. 
The resulting longitudinal noise, which modulates 
the energy levels, and transverse noise, which induces 
transitions, may therefore originate from common bath 
degrees of freedom and exhibit nonvanishing cross 
correlations. When these correlations are appreciable, the 
corresponding cross-spectral densities contribute to the 
reduced dynamics, rendering an independent-channel 
description inadequate. 
This motivates a unified theoretical treatment 
in which both noise components and their mutual 
correlations are retained.\\

Such multichannel-correlated-noise noise can 
arise in several 
experimentally relevant solid-state platforms. 
In superconducting qubits, fluctuations in charge, 
magnetic flux, or junction critical current modify parameters of the qubit Hamiltonian and, depending 
on the operating point and energy eigenbasis, 
can contribute to both dephasing and relaxation \cite{Makhlin2001,Ithier2005,Paladino2014}.Related effects may occur in semiconductor spin qubits and defect-based spin systems subject to fluctuating nuclear or electronic fields \cite{Witzel2006}, as well as in systems coupled to phononic or electromagnetic environments whose mode structure is sampled by more than one system observable \cite{Krummheuer2002,McCutcheonNazir2010,NazirSchaller2018}.  Correlations generated by common environments are known to modify decoherence and dissipative dynamics in ways that cannot, in general, be reproduced by independent noise models \cite{Palma1996,FicekTanas2002,Jeske2013}.\\

The spin--boson model provides a paradigmatic framework for studying the reduced dynamics of a two-level system coupled to environmental harmonic modes \cite{Leggett1987,Weiss2012,Breuer2002}. Within this framework, the statistical properties of a multichannel environment are naturally described by a matrix-valued spectral density $\mathbf{J}(\omega)$. Its diagonal elements specify the auto-spectral densities of the individual coupling channels, whereas its off-diagonal elements describe their cross-spectral correlations in the frequency domain. The microscopic construction of $\mathbf{J}(\omega)$ from the bath coupling amplitudes requires it to be Hermitian and positive semidefinite, thereby constraining the magnitude of the cross spectrum relative to the corresponding diagonal spectra. The cross-spectral density may nevertheless remain complex, with its magnitude determining the correlation strength and its phase encoding relative temporal or phase relationships between the environmental fluctuations.\\

The dynamical consequences of such cross-spectral correlations are particularly relevant when the two bath channels couple to non-commuting system observables. In this case, the cross terms connect processes conventionally identified with dephasing and relaxation and may modify both the damping and coherent evolution of the system. A treatment based solely on separate dephasing and relaxation rates is therefore insufficient whenever these correlations are appreciable. Moreover, structured or finite-bandwidth environments retain memory over non-negligible timescales, making a finite-time description preferable to a purely Markovian treatment \cite{Garraway1997,Mazzola2009,Pleasance2020}. The time-convolutionless projection-operator method provides a convenient time-local formulation in which such memory effects are retained through explicitly time-dependent coefficients determined by finite-time integrals of the bath correlation functions \cite{Breuer2002,RivasHuelga2012}.\\

In the present work, we investigate a qubit coupled to a shared bosonic bath through the interaction
$H_{SB}=
\sigma_x\otimes B_x+\sigma_z\otimes B_z$
where the transverse and longitudinal coupling operators may probe the same environmental modes. The resulting mode-weighted overlap gives rise to the off-diagonal spectral densities $J_{xz}(\omega)$ and $J_{zx}(\omega)=J_{xz}^{*}(\omega)$. Within the second-order time-convolutionless (TCL2) approximation, we derive a closed time-local master equation for the reduced state $\rho(t)$.
The corresponding finite-time TCL2 dynamics of the Bloch-vector components of $\rho(t)$ is governed by a time-dependent matrix $\mathbf{M}(t)$ and an inhomogeneous vector contribution $\mathbf{K}(t)$, which consistently integrate dissipative deformation, nonunital drift, and bath-induced coherent renormalization into the reduced dynamics.
We obtain analytic expressions for  $\mathbf{M}(t)$ and  $\mathbf{K}(t)$ as frequency-domain integrals of the auto- and cross-spectral densities, weighted by thermal factors and TCL2 time-dependent kernel functions arising within the TCL2 formulation.\\

To describe a broad class of correlated environments, we parametrize the cross-spectral density as
\begin{eqnarray}
J_{xz}(\omega)=
\sqrt{J_{xx}(\omega)J_{zz}(\omega)}\gamma(\omega) 
\end{eqnarray}
where the complex correlation coefficient satisfies $|\gamma(\omega)|\leqslant1$.
We adopt (fixed) Ohmic-type forms for the auto-spectral densities $J_{xx}(\omega)$ and $J_{zz}(\omega)$,
while the frequency-dependent cross correlations are
characterized by their strength, spectral bandwidth, relative delay, and phase offset involved in the parametrization of $\gamma(\omega)$.  This parametrization enables a direct comparison between correlated and uncorrelated dynamics while leaving the individual dephasing and relaxation spectra unchanged.\\

The numerical implementation is first validated in two standard limiting cases. In the purely longitudinal limit, the numerical Bloch-vector dynamics reproduces the exact solution of the pure-dephasing spin--boson model. In the purely transverse limit, it recovers the established qualitative dependence of the dissipative spin--boson dynamics on the Ohmicity of the bath spectrum \cite{Barr2024}. We then show that the cross-spectral terms couple the otherwise distinct dephasing and relaxation sectors, producing dynamics that cannot be obtained by simply adding the effects of the two independent channels. In particular, the correlations generate a non-monotonic modulation of the population relaxation and can induce a transient revival of the $\ell_1$-norm coherence following its initial decay. The magnitude and temporal structure of these effects depend sensitively on the correlation strength, bandwidth, delay, and phase.\\

These results show that cross correlations do not uniformly suppress environmental decoherence. Instead, they redistribute coherence loss and population relaxation in time, producing alternating intervals of enhanced and suppressed decoherence relative to the uncorrelated dynamics. Such behavior may provide finite temporal windows of improved coherence preservation or reduced relaxation within the regime of validity of the weak-coupling approximation. We emphasize, however, that the revival of a particular coherence measure does not by itself establish information backflow, genuine non-Markovianity according to a specific measure, or an operational quantum advantage; these properties require separate dynamical and task-dependent analyses.\\

 The rest of the article is organized as follows. In Sec.~\ref{sec:qb}, we introduce the microscopic system--bath model and characterize the correlated environmental fluctuations in terms of a Hermitian, positive-semidefinite spectral-density matrix. In Sec.~\ref{eq:tcl2}, we derive the general time-convolutionless master equation to second order in the system--bath coupling. Sec.~\ref{sec:qubittcl2} specializes this formalism to a qubit subject to correlated longitudinal and transverse noise and obtains the corresponding Bloch-vector equations in terms of the time-dependent matrix $\mathbf{M}(t)$ and the inhomogeneous vector $\mathbf{K}(t)$. In Sec.~\ref{sec:sd}, we introduce physically motivated parametrizations of the auto- and cross-spectral densities. Sec.~\ref{sec:nu} describes the numerical implementation and validates it against the exactly solvable pure-dephasing limit and the established behavior of the transverse-coupling spin--boson model. In Sec.~\ref{sec:cross_effects}, we analyze the effects of cross-spectral correlations on the qubit coherence and population dynamics by comparing correlated and uncorrelated environments. Finally in Sec.~\ref{sec:tr}, we discuss the operational significance and potential technological relevance of the correlation-induced coherence-revival and relaxation-suppression windows for quantum-state storage, control, sensing, and readout, while emphasizing the need for task-specific performance measures.

\section{Microscopic Model and Spectral Characterization of Correlated Baths}
\label{sec:qb} 
We now specify the microscopic model used to describe the correlated
longitudinal and transverse fluctuations discussed in
Sec.~\ref{sec:intro}. The joint system--bath Hilbert space is
$\mathcal{H}_S\otimes\mathcal{H}_B$, and the total Hamiltonian is
written as
\begin{eqnarray}
H
&=&
H_S\otimes\mathbb{I}_B
+
\mathbb{I}_S\otimes H_B
+
\gamma H_{SB},
\label{eq:totH}
\end{eqnarray}
where $H_S$ and $H_B$ are the bare system and bath Hamiltonians,
respectively, and $H_{SB}$ describes the system--bath interaction.
The operators $\mathbb{I}_S$ and $\mathbb{I}_B$ denote the identity
operators on $\mathcal{H}_S$ and $\mathcal{H}_B$, respectively. The
dimensionless parameter $\gamma$ controls the overall coupling strength,
with $H_{SB}$ taken to have dimensions of energy, and is retained as a
bookkeeping parameter for the perturbative expansion. It may be absorbed
into the definition of $H_{SB}$ after the perturbative order has been
identified.
The bath is modeled as a collection of independent bosonic modes.
Setting $\hbar=1$ and omitting  zero-point-energy contribution,
bare Hamiltonian of the bath can be written as
\begin{eqnarray}
H_B
&=&
\sum_k \omega_k b_k^\dagger b_k,
\label{eq:bathH}
\end{eqnarray}
where $b_k^\dagger$ and $b_k$ are the creation and annihilation
operators of the $k$-th oscillator mode of frequency $\omega_k$ and 
satisfy the canonical commutation relations
$
[b_k,b_{k'}^\dagger] = 
\delta_{kk'}$ and $[b_k,b_{k'}]=
[b_k^\dagger,b_{k'}^\dagger]=0$.
Each mode possesses an unbounded Fock ladder with occupation number
$n_k=0,1,2,\cdots$.   
The bath is initially assumed to be in 
thermal equilibrium at
temperature $T$ and is described by the Gibbs state
$\rho_B = 
\frac{e^{-\beta H_B}}{Z_B}$,
where $Z_B = \operatorname{Tr}_B\!\left(e^{-\beta H_B}\right)$ is the bath partition function
and $\beta = \frac{1}{k_B T}$ is the inverse bath
temperature.
Since
$[\rho_B,H_B]=0$, the bath state is stationary under the free evolution
generated by $H_B$. \\
%We additionally assume an initially factorized total state:
%$\rho_{\mathrm{tot}}(0)=\rho_S(0)\otimes\rho_B$.\\

The system--bath interaction is taken to be
\begin{eqnarray}
H_{SB}
&=&
\sum_\alpha S_\alpha\otimes B_\alpha,
\label{eq:HSB_model}
\end{eqnarray}
where $S_\alpha$ and $B_\alpha$ act on 
Hilbert spaces $\mathcal{H}_S$ and $\mathcal{H}_B$, respectively, and $\alpha$ labels the coupling channels. The bath operators characterize the environmental fluctuations, while the system operators specify how these 
bath-fluctuations couple to the system. Operators commuting with $H_S$ describe longitudinal coupling, whereas operators with nonvanishing matrix elements between mutually
orthogonal energy eigenstates describe transverse coupling.
Although Hermiticity of $H_{SB}$ 
does not imply that the operators appearing in an arbitrary tensor-product decomposition are individually Hermitian, any Hermitian interaction Hamiltonian can be decomposed into a sum of tensor products of Hermitian system and bath operators. We therefore adopt a representation in which
$S_\alpha^\dagger=S_\alpha$ and 
$B_\alpha^\dagger=B_\alpha$. \\
 
We consider bath operators that are linear in the bosonic creation and annihilation operators, 
\begin{eqnarray}
B_\alpha
&=&
\sum_k
\left(
g_{k\alpha}b_k
+
g_{k\alpha}^{*}b_k^\dagger
\right),
\label{eq:bath0}
\end{eqnarray}
where $g_{k\alpha}$  denotes the complex coupling coefficient between the $k$-th bath mode and the system operator associated with the interaction channel
$\alpha$. 
%The combination in Eq.~\eqref{eq:bath0} is Hermitian for arbitrary complex $g_{k\alpha}$. 
Such linear coupling arises naturally when a system
interacts with small fluctuations of phononic, electromagnetic, or other
approximately harmonic environmental fields.  
Different system coupling channels may couple to
the same set of bath modes, such that the corresponding coupling vectors, $\{g_{k\alpha}\}$ and $\{g_{k\beta}\}$, need not be orthogonal for $\alpha\neq\beta$. A nonvanishing mode-weighted overlap between these vectors gives rise to off-diagonal bath correlation functions, $C_{\alpha\beta}(t)$, and the associated cross-spectral densities. The
coupling distinct system channels to common bath modes therefore provides a possible microscopic origin for correlations between longitudinal and transverse environmental fluctuations.\\

To characterize the statistical properties of the bath,
we work in the Heisenberg picture, where the bath state remains stationary, $\rho_B$, and the bath operators evolve under the free bath Hamiltonian $H_B$: 
\begin{eqnarray}
B_\alpha(t)  &=& e^{iH_B t} B_\alpha e^{-iH_B t}
= \sum_k g_{k\alpha} b_k e^{-i\omega_k t}
+ g_{k\alpha}^\star b_k^\dagger e^{i\omega_k t}
\label{eq:bath1}
\end{eqnarray}
The statistical properties of the environment are encoded in the forward and backward temporal correlators 
\begin{eqnarray}
C_{\alpha\beta}(t)  
= \trd{B}{\Big[ B_\alpha(t) B_\beta(0)\rho_B\Big]} \quad \mbox{and} \quad
C_{\alpha\beta}(-t)  
= \trd{B}{\Big[ B_\alpha(0) B_\beta(t)\rho_B\Big]} \,\label{eq:bath2}
\end{eqnarray}
where the expectation values of the product 
of bath operators at two different times
are taken with 
respect to the thermal
Gibbs state $\rho_B$ of the bath.  Since the bath operators are Hermitian, the correlation functions satisfy $C_{\alpha\beta}(t) = \big[C_{\beta\alpha}(-t)\big]^\ast$.
and using Eq.~(\ref{eq:bath1}) one obtains
\begin{eqnarray}
C_{\alpha\beta}(t)  &=& \av{B_\alpha(t)B_\beta(0)}
=
\sum_k \Big[ g_{k\alpha}g_{k\beta}^\star \ (n_k+1) e^{-i\omega_k t}
+ g_{k\alpha}^\star g_{k\beta}\ n_k e^{-i\omega_k t} \Big]
\label{eq:cfull1}
\end{eqnarray}
where we have used the canonical commutation relations
$\com{b_k}{b_{k'}^\dagger}=\delta_{kk'}$ together with the thermal
expectation values
$\av{b_k^\dagger b_k}=n_k$ and $\av{b_k b_k^\dagger}=n_k+1$.
The equilibrium occupation number $n_k$
of the $k$-th bosonic mode is given by the Bose--Einstein distribution 
$n_k \equiv n(\omega_k) = (e^{\beta\omega_k}-1)^{-1}$.\\

Since the equilibrium bath state is stationary, the correlation functions depend only on the time difference. This time-translation invariance motivates the frequency-domain representation
\begin{eqnarray}
S_{\alpha\beta}(\omega)
&\equiv&
\int_{-\infty}^{\infty} dt\,
e^{i\omega t}C_{\alpha\beta}(t)
\nonumber\\
&=&
2\pi\sum_k
\left[
g_{k\alpha}g_{k\beta}^{*}(n_k+1)\delta(\omega-\omega_k)
+
g_{k\alpha}^{*}g_{k\beta}n_k\delta(\omega+\omega_k)
\right],
\label{eq:bath3}
\end{eqnarray}
where \(n_k\equiv n(\omega_k)\), and we have used
$\int_{-\infty}^{\infty}dt\,
e^{i(\omega\pm\omega_k)t}
=
2\pi\delta(\omega\pm\omega_k)$. 
The noise spectra may be written separately for positive and negative $\omega$ as
\begin{eqnarray}
S_{\alpha\beta}(\omega)
=
\begin{cases}
J_{\alpha\beta}(\omega)
\bigl[n(\omega)+1\bigr],
& \omega \geqslant ,\\[1mm]
J_{\alpha\beta}^{*}(|\omega|)
n(|\omega|),
& \omega<0,
\end{cases}
\label{eq:bath4}
\end{eqnarray}
where the matrix-valued bath spectral density is defined for positive
frequencies by
\begin{eqnarray}
J_{\alpha\beta}(\omega)
&\equiv&
2\pi\sum_k
g_{k\alpha}g_{k\beta}^{*}
\delta(\omega-\omega_k),
\qquad \omega>0\,,
\label{eq:bath5}
\end{eqnarray}
which satisfies the Hermiticity condition
\begin{equation}
J_{\alpha\beta}(\omega)
=
J_{\beta\alpha}^{*}(\omega).
\label{eq:bath5a}
\end{equation}
%
%The diagonal elements \(J_{\alpha\alpha}(\omega)\) describe the
%individual noise channels, whereas the off-diagonal elements
%\(J_{\alpha\beta}(\omega)\), with \(\alpha\neq\beta\), characterize
%correlations between distinct coupling channels.

Using the inverse Fourier transform, 
the temporal 
bath correlation function can be expressed  
 in terms of the spectral density 
 $J_{\alpha\beta}(\omega)$ as
\begin{eqnarray}
C_{\alpha\beta}(t)
&=&
\frac{1}{2\pi} \int_{-\infty}^{\infty} d\omega e^{-i\omega t}S_{\alpha\beta}(t)  =
\int_{0}^{\infty} d\omega 
\Big[J_{\alpha\beta}(\omega) (n(\omega)+1) e^{-i\omega t} 
+J_{\alpha\beta}^\ast(\omega)n(\omega) e^{i\omega t} \Big]\nonumber\\
\label{eq:bath6}
\end{eqnarray}
The matrix $\mathbf{J}(\omega)$, with elements $J_{\alpha\beta}(\omega)$  is referred to as the
spectral density matrix associated with the bath. The  condition
\eqref{eq:bath5a} ensures that 
$J(\omega)$ is Hermitian. Moreover, from
its microscopic definition  (Eq.~\eqref{eq:bath5}) 
it follows that the  matrix $\mathbf{J}(\omega)$ is  
positive semidefinite for each $\omega>0$, 
since it can be written as a $\delta(\omega-\omega_k)$-weighted sum of
outer products as $\mathbf{J}(\omega) 
= 2\pi \sum_k\delta(\omega-\omega_k)  \mathbf{g}_k\mathbf{g}_k^\dagger$ where $\mathbf{g}_k
= (g_{k1} \ g_{k2} \cdots)^T$ is a column vector involving the couplings
$g_{k\alpha}$. The Positive semidefiniteness guarantees nonnegative noise power for any linear combination of coupling channels and bounds the cross spectra by the corresponding diagonal components.\\
 
The diagonal elements 
$J_{\alpha\alpha}(\omega)
=
2\pi\sum_k |g_{k\alpha}|^2\delta(\omega-\omega_k)$
describe the
individual noise channels $\alpha$ and
measure the coupling-weighted density of bath modes   thereby characterizing the noise strength acting through channel $\alpha$ at frequency $\omega$. By contrast, the off-diagonal elements $J_{\alpha\beta}(\omega)$, with $\alpha\neq\beta$, 
characterize and 
quantify cross-spectral correlations between distinct coupling channels $\alpha$ and $\beta$. They are nonzero when the system operators $S_\alpha$ and $S_\beta$ couple to common bath modes and encode the resulting correlated contribution to the reduced-system dynamics.

\section{Time-Convolutionless Master Equation at Second Order} 
\label{eq:tcl2} 
The joint system--bath state obeys the von Neumann equation
\begin{eqnarray}
\frac{d\rho_{\mathrm{tot}}(t)}{dt}
&=&
-i\big[H,\rho_{\mathrm{tot}}(t)\big],
\label{eq:rd1}
\end{eqnarray}
where the Hamiltonian $H$ is decomposed as
$H = H_0 + \gamma H_{SB}$, with
 $H_0 \equiv H_S \otimes \mathbb{I}_S + \mathbb{I}_B \otimes H_B$.
 Although Eq.~(\ref{eq:rd1}) 
 governs the exact composite dynamics, a closed, time-local equation for the reduced state,  
 $\rho(t) = \operatorname{Tr}_B\left[\rho_{\mathrm{tot}}(t)\right]$ may be obtained perturbatively using the time-convolutionless projection-operator technique.
 It is 
 convenient to work in the interaction picture with respect
to the free Hamiltonian $H_0$, which separates the
trivial free evolution of the system and bath from the dynamics
induced by the system--bath interaction yielding an
equation of motion which 
is governed solely by the interaction Hamiltonian as,
\begin{eqnarray}
\frac{d \tilde{\rho}_{\rm tot}}{dt}
&=& -i\gamma \big[ \tilde{H}_{SB}(t), \tilde{\rho}_{\rm tot}(t) \big].
\label{eq:rd5}
\end{eqnarray}
where, $\tilde{\rho}_{\rm tot}(t) = e^{i H_0 t}\, \rho_{\rm tot}(t)\, e^{-i H_0 t}$  and $\tilde{H}_{SB}(t)
= e^{i H_0 t}\, H_{SB}\, e^{-i H_0 t}$ are the  total 
density operator $\tilde{\rho}_{\rm tot}(t)$ 
and the interaction Hamiltonian $\tilde{H}_{SB}(t)$ in 
interaction picture.  Using  the form in Eq.\ \eqref{eq:HSB_model}, $\tilde{H}_{SB}(t)$ can be expressed 
as
\begin{eqnarray}
\tilde{H}_{SB}(t) = \sum_\alpha \tilde S_\alpha(t) \otimes  \tilde B_\alpha(t),
\label{eq:rd3}
\end{eqnarray}
where 
$\tilde S_\alpha(t) = e^{i H_S t} S_\alpha e^{-i H_S t}$ and  $\tilde B_\alpha(t) = e^{i H_B t} B_\alpha e^{-i H_B t}$
denote the free Heisenberg evolution of the system and bath operators on 
their respective Hilbert spaces.  A formal integration
of Eq.~(\ref{eq:rd5}), followed by substitution back into the equation of motion,  
yields the
 iterated integral form 
\begin{eqnarray}
\frac{d \tilde{\rho}_{\rm tot}}{dt}
&=& -i \gamma\bigcom{\tilde{H}_{SB}(t)}{\tilde{\rho}_{\rm tot}(0)}
- \gamma^2 \int_0^t ds \bigcom{\tilde{H}_{SB}(t)}{\com{\tilde{H}_{SB}(s)}{\tilde{\rho}_{\rm tot}(s)}}\,,
\label{eq:rd7}
\end{eqnarray}
which is an exact identity and involves no approximation.  
In the weak-coupling regime, when the interaction-picture   state $\widetilde{\rho}_{\rm tot}$ changes with
time only at order $\gamma$,   we can substitute 
$\widetilde{\rho}_{\rm tot}(s)
=
\widetilde{\rho}_{\rm tot}(t)
+
\mathcal{O}(\gamma)$
 into the second-order term of Eq.~(\ref{eq:rd7}).
 This   generates corrections of order $\mathcal{O}(\gamma^3)$, which are consistently neglected  within the
second-order time-convolutionless (TCL2) approximation.
Taking the partial trace over the bath degrees of freedom yields the TCL2 equation of motion for the interaction-picture reduced density operator  
$\widetilde{\rho}$ as
\begin{eqnarray}
\frac{d\widetilde{\rho}(t)}{dt}
&=&
-i\gamma\,\trd{B}{\bigcom{\widetilde{H}_{SB}(t)}
{\widetilde{\rho}_{\rm tot}(0)}}
-\gamma^2\int_0^t ds\,
\trd{B}{\bigcom{\widetilde{H}_{SB}(t)}
{\com{\widetilde{H}_{SB}(s)}
{\widetilde{\rho}_{\rm tot}(t)}}}.
\label{eq:rd8}
\end{eqnarray}
%
%.
Assuming an initially factorized system--bath state
$\tilde{\rho}_{\rm tot}(0)=\tilde{\rho}(0)\otimes\rho_B$, 
and that the
bath operators have vanishing expectation 
values in the stationary bath
state $\rho_B$, i.e., $\mathrm{Tr}_B[B_\alpha(t)\rho_B]=0$, 
a condition
typically satisfied for bosonic environments 
in thermal equilibrium,
the term linear in $\gamma$ in Eq.~(\ref{eq:rd8}) 
vanishes.   
In the weak-coupling regime, system--bath correlations remain perturbatively small and are generated only at order $\gamma$. Accordingly, within the Born approximation one writes $\widetilde{\rho}_{\rm tot}(t)
=
\widetilde{\rho}(t)\otimes\rho_B
+
\mathcal{O}(\gamma)$,
with the bath state $\rho_B$ taken to be stationary. Substituting this form into the second-order kernel of Eq.~\eqref{eq:rd8} produces corrections of order $\mathcal{O}(\gamma^3)$, which lie beyond the accuracy of the TCL2 approximation. Under these assumptions, and performing the change of integration
variable $s\rightarrow t-s$, Eq.~(\ref{eq:rd8}) reduces to
\begin{eqnarray}
\frac{d \tilde{\rho}(t)}{dt}
&=&  
- \int_0^t ds \ \mathrm{Tr}_B 
\bigcom{\tilde{H}_{SB}(t)}
{\com{\tilde{H}_{SB}(t-s)}{\tilde{\rho}(t)\otimes\rho_B}}\,.
\label{eq:rd10}
\end{eqnarray}
where  the overall coupling strength $\gamma$ has been absorbed into the definition of the interaction Hamiltonian $\widetilde{H}_{SB}$. 
Using the form \eqref{eq:rd3} we can evaluate 
the commutators within memory kernel of 
Eq.\ \eqref{eq:rd10} to obtain
the resulting master equation
\begin{eqnarray}
\frac{d \tilde{\rho}(t)}{dt}
&=&  
- \int_0^t ds 
\sum_{\alpha,\beta} \Big( \com{\tilde S_\alpha(t)}{ \tilde S_\beta(t-s)\tilde{\rho}(t)}C_{\alpha\beta}(s)  -   \com{\tilde S_\alpha(t)}{ \tilde{\rho}(t)\tilde S_\beta(t-s)}C_{\beta\alpha}(-s)
\Big) \nonumber\\
\label{eq:rd11}
\end{eqnarray}
which is time-local in the sense that the derivative
$d\tilde{\rho}(t)/dt$ depends only on the 
instantaneous reduced state
$\tilde{\rho}(t)$. 
Nevertheless, the resulting dynamics is generally
non-Markovian as memory effects persist through the explicit time
dependence of the generator, which involves integrals over bath
correlation functions extending from the initial time up to the
observation time $t$.
Using   $\tilde S_\alpha(t) = e^{i H_S t} S_\alpha e^{-i H_S t}$ and 
$\tilde \rho(t) = e^{i H_S t} \rho(t) e^{-i H_S t}$
Eq.~\eqref{eq:rd11} can be transformed back to the Schr\"odinger picture, yielding the TCL2 master equation 
\begin{eqnarray}
 \dv{\rho(t)}{t}  
 &=&
 - i\com{H_S}{\rho(t)} -   \int_0^t ds 
\sum_{\alpha,\beta} \Big( \com{S_\alpha}{ \tilde S_\beta(-s) \rho(t) }C_{\alpha\beta}(s)  -   \com{S_\alpha}{  \rho (t)\tilde S_\beta(-s)}C_{\beta\alpha}(-s)
\Big)\nonumber\\  \label{eq:rd12} 
\end{eqnarray}
For an arbitrary system observable $A$ that is time independent in the
Schr\"odinger picture, evolution of the 
expectation value $\av{A}_t \equiv \trd{S}{[A\rho(t)]}$
follows directly from the TCL2 master equation 
\eqref{eq:rd12} as 
\begin{eqnarray}
 \dv{}{t}   \av{A}_t
 &=&
 - i \ \trd{S}{\Big(A\com{H_S}{\rho(t)}\Big)} \nonumber\\
 && -   \int_0^t ds 
 \sum_{\alpha,\beta} \Big[ 
\trd{S}{\Big( A\com{S_\alpha}{ \tilde S_\beta(-s) \rho(t) } \Big)}
 C_{\alpha\beta}(s)  -   
\trd{S}{\Big( A\com{S_\alpha}{  \rho (t)\tilde S_\beta(-s)}\Big)}
 C_{\alpha\beta}^\ast(s)
\Big]  \nonumber\\
\label{eq:rd13}
\end{eqnarray}
where we have used the relation
$C_{\beta\alpha}(-s)=C_{\alpha\beta}^{*}(s)$.

\section{TCL2 Dynamics of a Qubit under Correlated Longitudinal and Transverse Noise}
\label{sec:qubittcl2}
We now apply the general TCL2 formalism developed above to a two-level system subject to correlated longitudinal and transverse environmental fluctuations. The qubit is coupled to a bosonic bath through two distinct interaction channels, which give rise to relaxation, pure dephasing, and cross-correlation effects. Our aim is to derive the resulting time-local equations governing the evolution of the qubit Bloch-vector components. \\

We consider a single-qubit Hamiltonian of the form
\begin{eqnarray}
H_S
&=&
-\frac{\omega_0}{2}\sigma_z .
\end{eqnarray}
The minus sign fixes the basis convention  that,
the $\sigma_z$ eigenstate $\ket{0}$  
with eigenvalue $+1$ is the ground state with energy $-\omega_0/2$, whereas $\ket{1}$ with eigenvalue $-1$ is the excited state with energy $+\omega_0/2$.
We consider two distinct system--bath interaction channels described by
\begin{eqnarray}
H_{SB}
&=&
\sigma_x\otimes B_x+\sigma_z\otimes B_z ,
\end{eqnarray}
where $\sigma_\alpha$ ($\alpha=x,z$) are Pauli operators acting on the qubit, and $B_x$ and $B_z$ denote the bath operators associated with the corresponding coupling channels. Since $[H_S,\sigma_x]\neq0$, the transverse coupling $\sigma_x\otimes B_x$ induces transitions between the energy eigenstates and therefore gives rise to energy relaxation through the exchange of quanta with the environment. On the contrary, since $[H_S,\sigma_z]=0$, the longitudinal coupling $\sigma_z\otimes B_z$ does not induce energy transitions and instead produces pure dephasing---the decay of coherence in the energy eigenbasis without an accompanying exchange of energy.  
The environmental fluctuations are characterized by the bath correlation functions
$C_{\alpha\beta}(t)=\langle B_\alpha(t)B_\beta(0)\rangle$, with
$\alpha,\beta\in{x,z}$, and by the associated spectral-density matrix
$J_{\alpha\beta}(\omega)$ in the frequency domain. The diagonal correlators,
$C_{xx}(t)$ and $C_{zz}(t)$, describe fluctuations within the transverse and longitudinal channels and are associated with relaxation and dephasing, respectively. If the bath operators $B_x$ and $B_z$ couple to common environmental modes, the off-diagonal correlators,
$C_{xz}(t)$ and $C_{zx}(t)$, may be nonzero and quantify cross correlations between the two noise channels. Correspondingly, the diagonal elements of $J_{\alpha\beta}(\omega)$ characterize the individual noise spectra, whereas its off-diagonal elements encode the cross-spectral correlations between $B_x$ and $B_z$. 
This simple two-channel model therefore provides a simple setting for examining how relaxation, pure dephasing, and correlations between the two environmental noise channels jointly influence the reduced dynamics within the TCL master-equation framework.\\

In particular, we investigate the TCL2 evolution of the 
expectation values of the three qubit spin components by 
setting $A=\sigma_i$,($i=x,y,z$).
For these choices, Eq.\ \eqref{eq:rd13} yields the corresponding equation of motion for each component:
 \begin{eqnarray}
 \dv{}{t}   \av{\sigma_i}_t
 &=&
 \frac{i\omega_0}{2}  \ \trd{S}{\Big(\sigma_i\com{\sigma_z}{\rho(t)}\Big)} \nonumber\\
 && 
 -   \int_0^t ds 
 \Big[ 
\trd{S}{\Big( \sigma_i\com{\sigma_x}{ \tilde \sigma_x(-s) \rho(t) } \Big)}
 C_{xx}(s)  -   
\trd{S}{\Big( \sigma_i\com{\sigma_x}{  \rho (t)\tilde \sigma_x(-s)}\Big)}
 C_{xx}^\ast(s)
\Big]  \nonumber\\
&& 
-   \int_0^t ds 
\Big[ 
\trd{S}{\Big( \sigma_i\com{\sigma_x}{  \sigma_z  \rho(t) } \Big)}
 C_{xz}(s)  -   
\trd{S}{\Big( \sigma_i\com{\sigma_x}{  \rho (t)  \sigma_z }\Big)}
 C_{xz}^\ast(s)
\Big]  \nonumber\\
&& 
-   \int_0^t ds 
   \Big[ 
\trd{S}{\Big( \sigma_i\com{\sigma_z}{ \tilde \sigma_x(-s) \rho(t) } \Big)}
 C_{zx}(s)  -   
\trd{S}{\Big( \sigma_i\com{\sigma_z}{  \rho (t)\tilde \sigma_x(-s)}\Big)}
 C_{zx}^\ast(s) 
\Big] \nonumber\\
 && 
 -   \int_0^t ds 
\Big[ 
\trd{S}{\Big( \sigma_i\com{\sigma_z}{  \sigma_z \rho(t) } \Big)}
 C_{zz}(s)  -   
\trd{S}{\Big( \sigma_i\com{\sigma_z}{  \rho (t)  \sigma_z }\Big)}
 C_{zz}^\ast(s)
\Big] 
\label{eq:rd14}
\end{eqnarray}
where the freely evolved system operators are given by
\begin{eqnarray}
\tilde\sigma_z(-s) &=& e^{i\omega_0   s\sigma_z/2} \sigma_z e^{-i\omega_0  s\sigma_z/2} = \sigma_z \label{eq:rd14a}\\
\tilde\sigma_x(-s)  &=& e^{i\omega_0   s\sigma_z/2} \sigma_x e^{-i\omega_0  s\sigma_z/2} = (\cos \omega_0 s) \sigma_x - (\sin \omega_0 s) \sigma_y  \label{eq:rd14b}
\end{eqnarray}
Substituting Eq.~\eqref{eq:rd14b} into Eq.~\eqref{eq:rd14} and evaluating the system traces, the equations of motion for the three Bloch-vector components can be written in the compact form
 \begin{eqnarray}
  \dv{}{t}
 \begin{bmatrix} \av{\sigma_x}_t \cr \av{\sigma_y}_t \cr \av{\sigma_z}_t  \end{bmatrix}
 &=& 
\mathbf{M}(t) 
 \begin{bmatrix} 
\av{\sigma_x}_t \cr \av{\sigma_y}_t \cr \av{\sigma_z}_t  \end{bmatrix}
 + \mathbf{K}(t) \,,
 \label{eq:rd15}
\end{eqnarray} 
where $\mathbf{M}(t)$  is a $3\times3$ matrix and 
$\mathbf{K}(t)$ is 3-component vector
with components   given by  
\begin{eqnarray}
&& M_{xx}(t) = -4 \int_0^t ds  \Big(\mbox{Re~}C_{zz}(s) \Big) \,,
\quad
M_{xy}(t) = \omega_0 \,,\quad
M_{xz}(t) = 4 \int_0^t ds \cos(\omega_0s)\Big(\mbox{Re~} C_{zx}(s)
 \Big)\nonumber\\
&&
M_{yx}(t) = -\omega_0 -   4\int_0^t ds\sin(\omega_0s) \Big(\mbox{Re~} C_{xx}(s)  \Big) \,,\quad \nonumber\\
&& M_{yy}(t) =  -4\int_0^tds \cos(\omega_0s) \Big(\mbox{Re~} C_{xx}(s)  \Big) - 4\int_0^t ds\Big(\mbox{Re~}  C_{zz}(s)\Big)\nonumber\\
&& M_{yz}(t) = -4\int_0^t ds\sin(\omega_0s) \Big(\mbox{Re~} C_{zx}(s)   \Big) \nonumber\\
&&
M_{zx}(t) = 4 \int_0^t ds   \Big(\mbox{Re~} C_{xz}(s)   \Big)\,,\quad  
M_{zy}(t) =  0  \,,\quad  
M_{zz}(t) = -4  \int_0^t ds   \cos(\omega_0s) \Big(\mbox{Re~}C_{xx}(s)  \Big) \nonumber\\ 
&&  K_x(t) = 4 \int_0^t ds\  \sin(\omega_0s) \Big(\mathrm{Im} \ C_{zx}(s) \Big)\nonumber\\
 && K_y(t) =  -4 \int_0^t ds\Big( \mathrm{Im}\ C_{xz}(s)  \Big)
+ 4\int_0^t ds\cos(\omega_0s) \Big( \mathrm{Im}\ C_{zx}(s) \Big)\nonumber\\  
&&  K_z(t) =  
-4 \int_0^t ds  \sin(\omega_0s)\Big(\mathrm{Im} \
 C_{xx}(s)\Big) 
\label{eq:rd16a}
\end{eqnarray} 
The bath correlation functions $C_{\alpha\beta}(s)$, with $\alpha,\beta\in{x,z}$, appearing in Eq.~\eqref{eq:rd16a} can be expressed in terms of the spectral densities $J_{\alpha\beta}(\omega)$ through Eq.~\eqref{eq:bath6},
where $J_{xx}(\omega)$ and $J_{zz}(\omega)$ are real, while the cross spectral densities are complex satisfying $J_{xz}(\omega)=J_{zx}^{*}(\omega)$. 
Using the Bose--Einstein occupation factor $n(\omega)=(e^{\beta\omega}-1)^{-1}$,
the elements of $\mathbf{M}(t)$ and $\mathbf{K}(t)$ may   be written as frequency integrals involving $J_{\alpha\beta}(\omega)$ as 
{\small
\begin{eqnarray} 
 M_{xx}(t) 
&=& 
-4 \int_{0}^{\infty} d\omega \;
J_{zz}(\omega)\,\coth\!\left(\frac{\beta\omega}{2}\right)
\phi_c(\omega,t)  \quad , \quad
M_{xy}(t)  
=
\omega_0 \nonumber\\
M_{xz}(t)  
&=&
4 \int_{0}^{\infty} d\omega 
\Big[
\coth\!\left(\frac{\beta\omega}{2}\right)\,
\Big(\mathrm{Re}\,J_{xz}(\omega)\Big) 
\phi_{cc}(\omega,t) 
- \Big(\mathrm{Im}\,J_{xz}(\omega)\Big) 
\phi_{cs}(\omega,t)  
\Big] \nonumber\\
M_{yx}(t)  
&=&
-\omega_0 -   4
\int_{0}^{\infty} d\omega \;
J_{xx}(\omega)\,\coth\!\left(\frac{\beta\omega}{2}\right)
\phi_{ss}(\omega,t) \nonumber\\
 M_{yy}(t)  
&=&
-4 \int_{0}^{\infty} d\omega \coth\left(\frac{\beta\omega}{2}\right)
\Bigg[ J_{xx}(\omega)\phi_{cc}(\omega,t) + J_{zz}(\omega) \phi_{c}(\omega,t)\Bigg]  \nonumber   \\
M_{yz}(t)  
&=&
-4\int_{0}^{\infty} d\omega 
\Bigg[\coth\left(\frac{\beta\omega}{2}\right)\Big(\mathrm{Re} \ J_{xz}(\omega)\Big) 
\phi_{sc}(\omega,t)
- \Big(\mathrm{Im} \ J_{xz}(\omega)\Big)
\phi_{ss}(\omega,t) 
\Bigg] \nonumber\\
M_{zx}(t)  
&=&
4\int_{0}^{\infty} d\omega 
 \coth\left(\frac{\beta\omega}{2}\right)
  \Big[ \big(\mathrm{Re} \ J_{xz}\big)
  \phi_c(\omega,t) +  \big(\mathrm{Im} \ J_{xz}\big)
   \phi_s(\omega,t) \Big] \nonumber\\
M_{zy}(t)  
&=&  
0 \quad , \quad 
M_{zz}(t)  
=  
  -4 
 \int_{0}^{\infty} d\omega J_{xx}(\omega)
\coth\left(\frac{\beta\omega}{2}\right) 
\phi_{cc}(\omega,t)
\label{eq:Mfinal}
\end{eqnarray}
}
and,
{\small
\begin{eqnarray}
 K_x(t)  
&=&
-4 
  \int_{0}^{\infty} d\omega \Bigg[ \big(\mathrm{Im} \ J_{xz}(\omega)\big)
\phi_{sc}(\omega,t)   + \big(\mathrm{Re} \ J_{xz}(\omega)\big)
\phi_{ss}(\omega,t)  \Bigg] \nonumber\\
K_y(t)  
&=&
-4 
\int_{0}^{\infty} d\omega \Big[ \big(\mathrm{Im} \ J_{xz}(\omega)\big)\phi_{c}(\omega,t)
  -  \big(\mathrm{Re} \ J_{xz}(\omega)\big)
\phi_{s}(\omega,t)    \Big]\nonumber\\
&&
\qquad - 4  
  \int_{0}^{\infty} d\omega \Big[ \big(\mathrm{Im} \ J_{xz}(\omega)\big)
 \phi_{cc}(\omega,t)   + \big(\mathrm{Re} \ J_{xz}(\omega)\big)
 \phi_{cs}(\omega,t)  \Big] \nonumber\\
K_z(t)  
&=&
 4  \int_{0}^{\infty} d\omega J_{xx}(\omega)
 \phi_{ss}(\omega,t)
 \label{eq:Kfinal}
\end{eqnarray}
} 
where  the functions $\phi_{j}(\omega,t)$'s are given by
\begin{eqnarray}
\phi_c(\omega,t) \equiv \int_0^t ds \cos(\omega s)&,&
\phi_s(\omega,t)  \equiv\int_0^t ds \sin(\omega s)\nonumber\\
\phi_{cc}(\omega,t) \equiv \int_0^t ds\cos(\omega_0 s)\cos(\omega s)&,& 
\phi_{cs}(\omega,t) \equiv \int_0^t ds\cos(\omega_0 s)\sin(\omega s)\nonumber\\ 
\phi_{sc}(\omega,t) \equiv \int_0^t ds \sin(\omega_0 s)\cos(\omega s) &,& 
\phi_{ss}(\omega,t) \equiv \int_0^t ds\sin(\omega_0 s)\sin(\omega s)  
\label{eq:phifunc}
\end{eqnarray}
Because the spectral density matrix $\mathbf{J}(\omega)$  
is Hermitian, for the two-component coupling (e.g., via $\sigma_x$ and $\sigma_z$) as considered here,
 it is fully characterized by the real
diagonal entries $J_{xx}(\omega)$, $J_{zz}(\omega)$, and the real and imaginary parts of the cross-spectral density $J_{xz}(\omega)$. 
Moreover, the positivity of   matrix $\mathbf{J}(\omega)$ imposes
the Cauchy--Schwarz inequality for bath correlation functions \cite{Breuer2002},
\begin{eqnarray}
J_{xx}(\omega) \geqslant 0\,,\quad
J_{yy}(\omega)  \geqslant 0 \,,\quad
|J_{xz}(\omega)|^2 \leqslant  J_{xx}(\omega) J_{yy}(\omega)\, 
\quad \mbox{(for all $\omega$)}\,,
\label{eq:Jcons}
\end{eqnarray}
which 
restrict the allowed correlations 
between different noise channels.
Elements of the spectral density matrix subject to the above 
constants,  completely determine the 
decoherence contributions encoded in  $\mathbf{M}(t)$ and $\mathbf{K}(t)$, which govern the time evolution of the expectation values $\av{\sigma_i}_t$.\\

For a chosen parametrization of the spectral-density matrix $\mathbf{J}(\omega)$, discussed in following Sec.~\ref{sec:sd}, the matrices $\mathbf{M}(t)$ and $\mathbf{K}(t)$ can be evaluated numerically. Eq.~(\ref{eq:rd15}) may then be solved for the expectation values $\av{\sigma_i}_t$ ($i=x,y,z$), subject to the boundary condition specified by the system's initial state $\rho(0)$. The resulting solution $(\av{\sigma_x}_t,\av{\sigma_y}_t),\av{\sigma_z}_t)$   constitute the components of the Bloch vector of the reduced state  
\begin{eqnarray}
\rho(t) &=& \frac12\Big[\mathbb{I} + \av{\sigma_x}_t\sigma_x 
+ \av{\sigma_y}_t\sigma_y + \av{\sigma_z}_t\sigma_z\Big]\,,
\label{eq:rhobloch}
\end{eqnarray}
at time $t$.
The coherence of the state in the eigenbasis of $\sigma_z$ is determined by the off-diagonal matrix elements, 
\begin{eqnarray} \rho_{01}(t) = \frac{1}{2} \left[ \langle\sigma_x\rangle_t -i\langle\sigma_y\rangle_t \right]
= \rho_{10}^\star(t) 
\end{eqnarray} 
Using the $l_1$-norm measure of coherence: 
$C(t) \equiv C\big[\rho(t)\big] = \sum_{i\neq j}|\rho_{ij}(t)|$, one obtains
\begin{eqnarray}
C(t)
&=& 
2|\rho_{01}(t)| = 
\sqrt{
\langle\sigma_x\rangle_t^2
+
\langle\sigma_y\rangle_t^2
}.
\label{eq:coherence}
\end{eqnarray}
The coherence $C(t)$ gives the magnitude of the transverse component of the Bloch vector $(\av{\sigma_x}_t,\av{\sigma_y}_y,\av{\sigma_z}_t)$ corresponding to $\rho(t)$, whereas $\av{\sigma_z}_t = \rho_{00}(t) - \rho_{11}(t)$ 
determines the population
imbalance in the chosen basis.\\

 We note that the TCL2 master equation in Eq.~\eqref{eq:rd12} is expressed as the sum of the bare Hamiltonian contribution and the complete finite-time TCL2 kernel. The resulting dynamics
is expressed in Bloch-vector representation by Eq.~\eqref{eq:rd15} in which the finite-time TCL2 generator is encoded in the time-dependent matrix $\mathbf{M}(t)$ and the  vector $\mathbf{K}(t)$.
 The symmetric part of $\mathbf{M}(t)$ governs the homogeneous nonunitary deformation of the Bloch vector, including dephasing, anisotropic damping, and possible transient amplification of its components. Although it may change the orientation of the Bloch vector through unequal attenuation or symmetric coupling between different components, it does not generate a rigid coherent rotation. The antisymmetric part of $\mathbf{M}(t)$ generates Bloch-vector rotations and therefore governs the effective coherent dynamics. It contains both the precession induced by the bare system Hamiltonian and the bath-induced coherent renormalization, the latter constituting the finite-time TCL2 analogue of the Lamb-shift Hamiltonian. The vector $\mathbf{K}(t)$,  describes the nonunital part of the reduced dynamics and produces a state-independent translation of the Bloch vector. 
In all numerical calculations presented here, the complete matrix $\mathbf{M}(t)$ and vector $\mathbf{K}(t)$ are retained, thereby consistently integrating the dissipative
deformation, nonunital drift, and Lamb-shift contributions into the dynamics rather than treating them separately.  \\

\section{On Parametric forms of Spectral Density Matrix}
\label{sec:sd}
The spectral density matrix provides a 
complete characterization of the
frequency-resolved correlations of bath operators.
For a two-component coupling a convenient parametrization,
consistent with the restrictions \eqref{eq:Jcons},
is obtained by introducing a
complex correlation coefficient $\gamma(\omega)$ defined as
\begin{eqnarray}
J_{xz}(\omega) &=&
\sqrt{J_{xx}(\omega) J_{zz}(\omega)} \, \gamma(\omega),
\label{eq:Jcross}
\end{eqnarray}
with $|\gamma(\omega)| \le 1$. The spectral density matrix can then be written as
\begin{eqnarray}
\mathbf{J}(\omega)  &=&
\begin{pmatrix}
J_{xx}(\omega) &
\sqrt{J_{xx}(\omega) J_{zz}(\omega)} \, \gamma(\omega) \\
\sqrt{J_{xx}(\omega) J_{zz}(\omega)} \, \gamma^*(\omega) &
J_{zz}(\omega)
\end{pmatrix}.
\label{eq:Jform}
\end{eqnarray}
The diagonal entries
$J_{xx}(\omega)$ and
$J_{zz}(\omega)$ characterize the auto-spectral densities -- the intrinsic noise strengths associated with
the fluctuations of the bath operators $B_x$ and $B_z$
coupled to the system observables
$\sigma_x$ and $\sigma_z$, respectively, 
and would fully describe the
environment in the absence of inter-channel correlations. 
The complex
coefficient $\gamma(\omega)$, 
  encodes the correlations
between these two noise channels. 
In particular, its magnitude
$|\gamma(\omega)|$ quantifies the degree of correlation, 
ranging from zero
for uncorrelated noise to unity for maximally correlated noise, while its
phase $\arg\gamma(\omega)$ 
captures the relative phase between the
fluctuations in the two channels.\\

A physically  motivated parametrization for the auto-spectral densities is the Ohmic-type form with an exponential cutoff,
\begin{eqnarray}
J_{\alpha\alpha}(\omega)
= \eta_\alpha\,\omega_{c\alpha}^{1-s_\alpha}\,\omega^{s_\alpha}
e^{-\omega/\omega_{c\alpha}}, \quad \alpha = x,z,
\label{eq:Jdiag}
\end{eqnarray}
where $\eta_\alpha$ denotes the system--bath coupling strength, $\omega_{c\alpha}$ is the cutoff frequency, and $s_\alpha$ controls the low-frequency scaling. The regimes $0<s_\alpha<1$, $s_\alpha=1$, and $s_\alpha>1$ correspond to sub-Ohmic, Ohmic, and super-Ohmic environments, respectively, characterized by enhanced, linear, and suppressed low-frequency spectral weight \cite{Leggett1987,Weiss2012,Breuer2002}. The exponential cutoff reflects the finite bandwidth set by microscopic energy scales and ensures regular behavior at high frequencies. This form provides an effective description of broadband reservoirs with weak spectral structure and short correlation times, and is widely employed in open quantum system models \cite{Caldeira1983,Clerk2010}, particularly in solid-state platforms such as superconducting qubits and semiconductor spin systems, as well as in quantum optical settings involving emitters coupled to weakly structured photonic environments away from sharp resonances. 
An alternative and equally important class of 
spectral densities arises in
structured environments possessing a dominant 
characteristic frequency, for
which a Lorentzian form provides an appropriate 
description. Such spectra
naturally occur when the system couples 
predominantly to a narrow band of
environmental modes, as in 
lossy cavity-QED systems and 
structured photonic
reservoirs~\cite{Garraway1997,Pleasance2020}, 
localized or weakly damped
molecular vibrations~\cite{Chin2013}, 
and non-Markovian reservoirs with finite
correlation times~\cite{Mazzola2009,Breuer2002}. 
In such cases, the
auto-spectral densities may be modeled as
\begin{eqnarray}
J_{\alpha\alpha}(\omega) 
= \frac{\lambda_\alpha \,\kappa_\alpha^2}
{(\omega - \omega_\alpha)^2 + \kappa_\alpha^2}, 
\qquad \alpha = x,z,
\label{eq:Jdiag1}
\end{eqnarray}
where $\omega_\alpha$ denotes the central (resonant) frequency of the bath, $\kappa_\alpha$ characterizes the spectral width (inverse correlation time), and $\lambda_\alpha$ sets the effective coupling strength. This form naturally emerges for environments with exponentially decaying correlations and finite memory timescales $\kappa_\alpha^{-1}$, and is widely used to model non-Markovian dynamics and resonance effects in open quantum systems \cite{Breuer2002,RivasHuelga2012,Garraway1997}. It is particularly relevant for structured reservoirs with well-defined spectral features, where deviations from the Markovian limit become significant.
In the present work, we adopt the Ohmic-type parametrization of the auto-spectral densities given in Eq.~\eqref{eq:Jdiag} to investigate the effects of cross-spectral correlations. \\\

 The remaining freedom in the parametrization lies in the choice of the complex correlation coefficient $\gamma(\omega)$, with $|\gamma(\omega)|\leqslant 1$, which determines the structure of the inter-channel correlations encoded in the complex cross-spectral density $J_{xz}(\omega)$. From a physical standpoint, a nonzero $\gamma(\omega)$ arises when the system operators $\sigma_x$ and $\sigma_z$ couple to common or partially overlapping environmental modes. The resulting cross correlations can produce interference between the corresponding noise channels \cite{Breuer2002,Clerk2010,Palma1996,Jeske2013}. Such situations can occur in structured reservoirs, electromagnetic environments in which distinct system couplings interact with overlapping field modes, and phononic environments in which different system degrees of freedom couple to overlapping vibrational mode distributions \cite{FicekTanas2002,Krummheuer2002,McCutcheonNazir2010}. 
The frequency dependence of $\gamma(\omega)$ also permits the description of environments in which correlations are confined to particular spectral regions, thereby capturing frequency-selective bath coherence and interference effects \cite{Leggett1987,Weiss2012,Clerk2010}.\\

One may adopt a  simple phenomenological choice 
for $\gamma(\omega) \equiv |\gamma(\omega)|e^{i\theta(\omega)}$ 
as an exponentially decaying
correlation magnitude,
\begin{eqnarray}
|\gamma(\omega)|
=
\gamma_0 e^{-\omega/\omega_c}   \qquad \mbox{and} \quad
\theta(\omega)=\omega\tau\ + \phi,,
\end{eqnarray}
where $0 \leqslant \gamma_0 \leqslant 1$ determines the strength of the cross-correlation,
$\omega_c$ defines the characteristic frequency scale beyond which the cross-correlations are exponentially suppressed.
The linear spectral phase $\theta(\omega)=\omega\tau$
represents a relative time delay $\tau$ between the two  correlated noise channels, while $\phi$ denotes a 
frequency-independent phase offset between them.
This parametrization represents an environment in which the two noise channels are strongly correlated at low frequencies, while their correlations gradually weaken as the frequency increases. Physically, this may occur when both system operators couple to the same low-energy bath modes but respond differently to higher-frequency modes. 
Such differences can arise because the operators probe distinct spatial regions of the environment, possess different wave-vector-dependent coupling strengths, or experience spatial averaging over rapidly varying bath fluctuations  \cite{Krummheuer2002,McCutcheonNazir2010,Jeske2013}. 
As a result, high-frequency modes may contribute less coherently to the two channels, leading to a progressive suppression of their cross correlations.

\section{Numerical Implementation and Validation in Limiting Cases}
\label{sec:nu}
In this section, we present numerical results for the reduced dynamics of a qubit within the second-order time-convolutionless (TCL2) approximation developed in Sec.~\ref{sec:qubittcl2}. 
For the different parametrizations of the spectral-density matrix $\mathbf{J}(\omega)$ introduced in Sec.\ \ref{sec:sd}, we evaluate the time-dependent Bloch-vector components and reconstruct the reduced state $\rho(t)$ together with its coherence $C(t)$.
The results are used to elucidate how inter-channel bath cross
correlations modify the evolution of the qubit state and, in particular, the decay, preservation, or possible revival of quantum coherence.\\

Throughout the numerical analysis, we employ natural units, $\hbar=k_B=1$, and use the qubit transition frequency $\omega_0$ as the reference energy scale, setting  the value $\omega_0=1$. Accordingly, all frequencies and energy scales, including  $\omega,\omega_{c\alpha}, \omega_c $ \textit{etc.}, are expressed in units of $\omega_0$. Time variables, including the evolution time ($t$) and the relative delay ($\tau$), are given in units of ($\omega_0^{-1}$). The coupling strengths $\eta_\alpha$, Ohmicity exponents  $s_\alpha$, correlation amplitude $\gamma_0$, and phase offsets ($\phi$) are dimensionless, with phases understood in radians.
The inverse temperature ($\beta$) is expressed in the same inverse-energy units. 
 In this convention, the dimensionless temperature is  $k_BT/(\hbar\omega_0)=1/(\beta\omega_0)$; hence, a choice $\beta=40$ corresponds to  $k_BT=0.025 \hbar\omega_0$, while $\beta=0.1$ corresponds to $k_BT=10\hbar\omega_0$. All parameter values quoted in the subsequent discussion are to be understood in these scaled units. \\

We assume that the qubit is initially prepared in the coherent superposition state $\ket{+}=\frac{\ket{0}+\ket{1}}{\sqrt{2}}$
with the corresponding density operator
\begin{eqnarray}
\rho(0)
&=&
\ket{+}\bra{+}
=
\frac{1}{2}
\begin{pmatrix}
1 & 1 \cr
1 & 1
\end{pmatrix}.
\label{eq:rho0}
\end{eqnarray}
The associated initial Bloch-vector components and initial coherence are:
\begin{eqnarray}
\av{\sigma_x}_{0}=1\,,\quad\av{\sigma_y}_{0}=0\,,\quad \av{\sigma_z}_{0}=0\,,\quad C(0)=1
\label{eq:dephini}
\end{eqnarray}
$\av{\sigma_x}_{0}=1$, $\av{\sigma_y}_{0}=0$, $\av{\sigma_z}_{0}=0$ 
while the initial coherence is  $C(0)=1$.\\

We numerically solve Eq.~\eqref{eq:rd15} for the Bloch-vector components $\av{\sigma_i}_t$, incorporating longitudinal dephasing, transverse relaxation, and their environmental cross correlations through the full spectral-density matrix $\mathbf{J}(\omega)$. Before addressing the general correlated case, we benchmark the longitudinal sector against the exact solution of the pure-dephasing spin-boson model. This comparison directly validates the dephasing contribution and the numerical routines used to evaluate the bath integrals and integrate the Bloch equations. As an independent check of the transverse sector, we further compare the amplitude-damping limit of our formulation with existing numerical results for the transverse-coupling spin-boson model.\\

\begin{figure}[t]
    \centering
    \begin{subfigure}[t]{0.48\columnwidth}
        \centering
        \includegraphics[width=\linewidth]{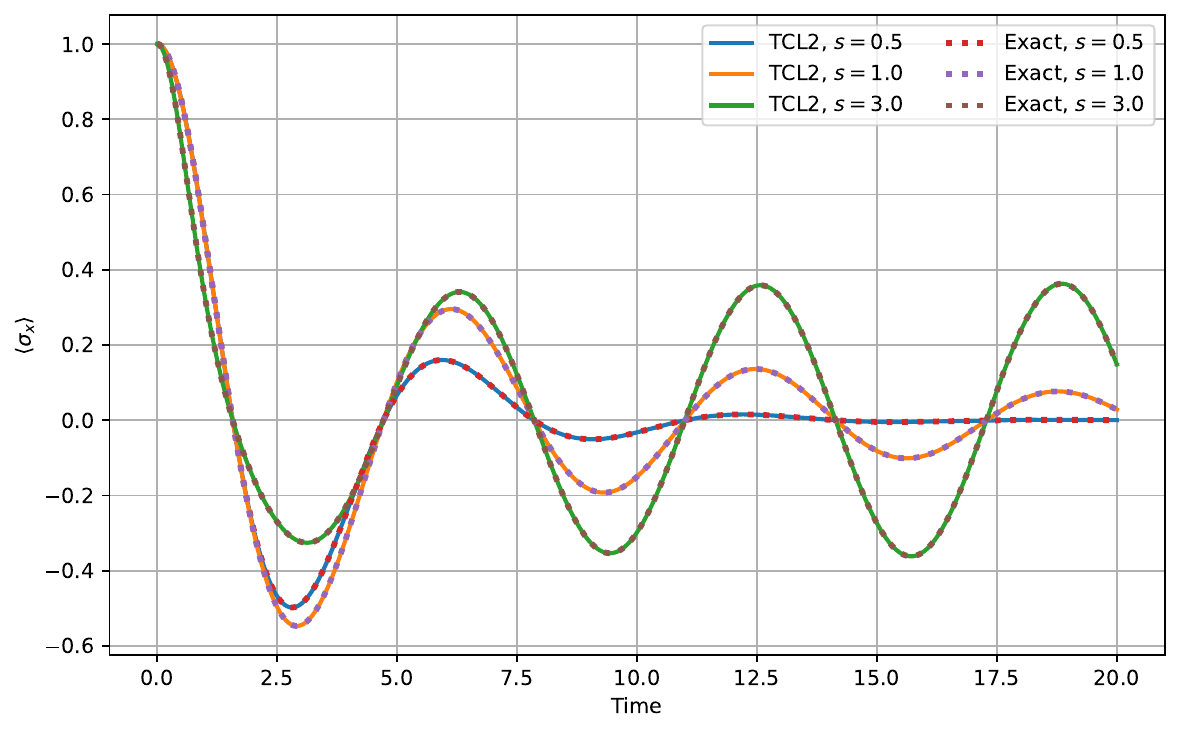}
%        \caption{Caption for the first figure.}
%        \label{fig:first}
    \end{subfigure}
    \hfill
    \begin{subfigure}[t]{0.48\columnwidth}
        \centering
        \includegraphics[width=\linewidth]{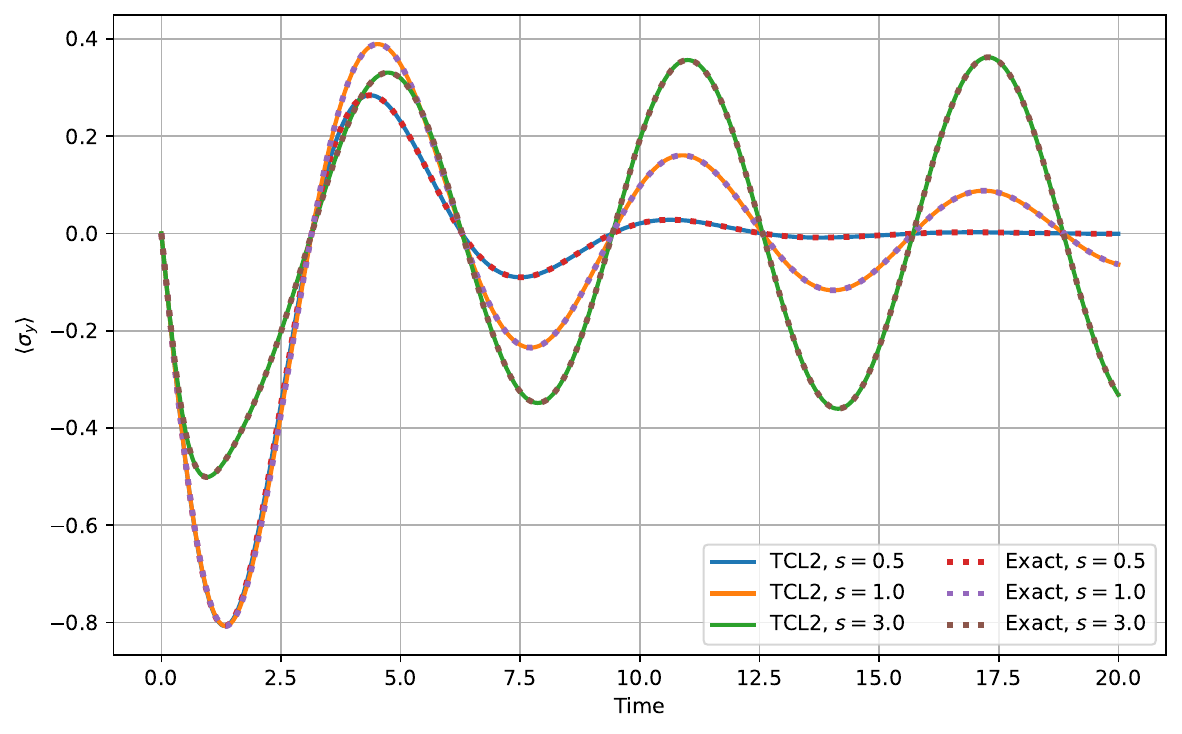}
%        \caption{Caption for the second figure.}
%       \label{fig:second}
    \end{subfigure}
\caption{Benchmarking against the exactly solvable pure-dephasing model: Temporal profile of $\av{\sigma_x}_t$ (left panel) and   $\av{\sigma_y}_t$ (right panel) for a purely dephasing environment. Solid curves show the results obtained from the numerical implementation after setting $J_{xx}(\omega)=J_{xz}(\omega)=J_{zx}(\omega)=0$, such that only $J_{zz}(\omega)$ contributes. The spectral-density $(J_{zz}(\omega))$ parameters are $\eta_z=0.25$, $\omega_{cz}=0.5$,  $s_z=0.5 \mbox{ and } 1.0$. Dotted curves show the corresponding exact pure-dephasing solutions given in Eq.~\eqref{eq:pure_dephasing_exact_bloch1} for the same parameter values.} 
    \label{fig:1}
\end{figure}

\textbf{Benchmark against Exact Pure Dephasing Dynamics:}
The pure-dephasing limit of the present two-channel model is obtained by
retaining only the longitudinal system--bath coupling. Accordingly, we set
\begin{eqnarray}
J_{xx}(\omega)=J_{xz}(\omega)=J_{zx}(\omega)=0,
\qquad
J_{zz}(\omega)\equiv J(\omega),
\label{eq:pure_dephasing_limit}
\end{eqnarray}
The TCL2 master equation\ \eqref{eq:rd12} 
in this pure dephasing limit reduces to 
\begin{eqnarray}
 \dv{\rho(t)}{t}  
 &=&- i\com{H_S}{\rho(t)} -     \frac{\gamma_{\rm pd}(t)}{2}  \Big(\rho(t)  - \sigma_z\rho(t)\sigma_z\Big)
\label{eq:pure_dephasing_limit1}
\end{eqnarray}
where,  
\begin{eqnarray}
\gamma_{\rm pd}(t)
&=& 
4 \int_0^\infty d\omega J(\omega)\coth\left(\frac{\beta\omega}{2}\right)\frac{\sin\omega t}{\omega}
\label{eq:gammapd}
\end{eqnarray}
With $H_S = -\omega_0\sigma_z/2$ and for any arbitrary initial state 
$\rho(0)$, the exact analytic solution of Eq.\ \eqref{eq:pure_dephasing_limit1} 
is given by
\begin{eqnarray}
\rho(t)
&=&
\begin{pmatrix}
\rho_{00}(0) & \rho_{01}(0)e^{i\omega_0 t}e^{-\Gamma_{\rm pd}(t)} \cr
\rho_{10}(0)e^{-i\omega_0 t}e^{-\Gamma_{\rm pd}(t)}&\rho_{11}(0)
\end{pmatrix}
\label{eq:dephsol}
\end{eqnarray}
where $\Gamma_{\rm pd}(t)$ is the integrated dephasing function
given by
\begin{eqnarray}
\Gamma_{\rm pd}(t) = \int_0^t ds \gamma_{\rm pd}(s)
= 4  \int_0^\infty d\omega J(\omega)\coth\left(\frac{\beta\omega}{2}\right)
\frac{1-\cos\omega t}{\omega^2}
\label{eq:Gammadeph}
\end{eqnarray}
Under pure
dephasing  the Bloch-vector components and the $\ell_1$-norm coherence
 evolve as
\begin{eqnarray}
\av{\sigma_x}_t
&=&
e^{-\Gamma_{\rm pd}(t)}
\left[
\av{\sigma_x}_0\cos(\omega_0 t)
+
\av{\sigma_y}_0\sin(\omega_0 t)
\right],
\nonumber\\
\av{\sigma_y}_t
&=&
e^{-\Gamma_{\rm pd}(t)}
\left[
\av{\sigma_y}_0\cos(\omega_0 t)
-
\av{\sigma_x}_0\sin(\omega_0 t)
\right],
\nonumber\\
\av{\sigma_z}_t
&=&
\av{\sigma_z}_0 \nonumber\\
C_{\rm pd}(t)
&=&
2\left|\rho_{01}(t)\right|
=
C_{\rm pd}(0)e^{-\Gamma_{\rm pd}(t)} .
\label{eq:pure_dephasing_exact_bloch}
\end{eqnarray}
Thus   the longitudinal component of Bloch vector remains constant, whereas the transverse
components undergo damped precession.  
For the initial state $\ket{+}$, with initial conditions as given in Eq.\ (\ref{eq:dephini}),
we have 
\begin{eqnarray}
\av{\sigma_x}_t
=
e^{-\Gamma_{\rm pd}(t)} 
 \cos(\omega_0 t)\,,\quad 
\av{\sigma_y}_t
=
- e^{-\Gamma_{\rm pd}(t)} 
 \sin(\omega_0 t)\, ,\quad  
\av{\sigma_z}_t
=
0\,, \quad 
C_{\rm pd}(t)  =
 e^{-\Gamma_{\rm pd}(t)} .
\label{eq:pure_dephasing_exact_bloch1}
\end{eqnarray}
For the spectral-density $J(\omega)$ corresponding to the  dephasing noise 
channel, we adopt the parametrization of
Eq.~\eqref{eq:Jdiag}, 
\begin{eqnarray}
 J(\omega) 
&=&
	\eta_z
	\omega_{cz}^{1-s_z}
	\omega^{s_z}
	\exp\left(
	-\frac{\omega}{\omega_{cz}}
	\right).
\label{eq:jzz}
\end{eqnarray}
Choosing different sets of values of the parameters
$(\eta_z,\omega_{cz},s_z)$ we use our code to numerically
compute $\av{\sigma_x}_t$ and $\av{\sigma_x}_t$   by setting $J_{xx}(\omega)=J_{xz}(\omega)=J_{zx}(\omega)=0$ and $J_{zz}(\omega)=J(\omega)$, and present the results in Fig.~\ref{fig:1} for $\eta_z=0.25,\omega_{cz}=0.5$ ,$s_z=(0.5 \mbox{ and } 1,0)$
with $\beta=40$. The numerical results agree with the exact pure-dephasing solutions in Eq.~\eqref{eq:pure_dephasing_exact_bloch1} to numerical precision. This benchmark validates the implementation of the longitudinal dephasing sector and the numerical routines used to evaluate the bath integrals and integrate the Bloch equations, thereby providing a reliable basis for the subsequent analysis of the full correlated two-channel dynamics.\\

 \begin{figure}[!h]
    \centering  
        \includegraphics[width=0.7\linewidth]{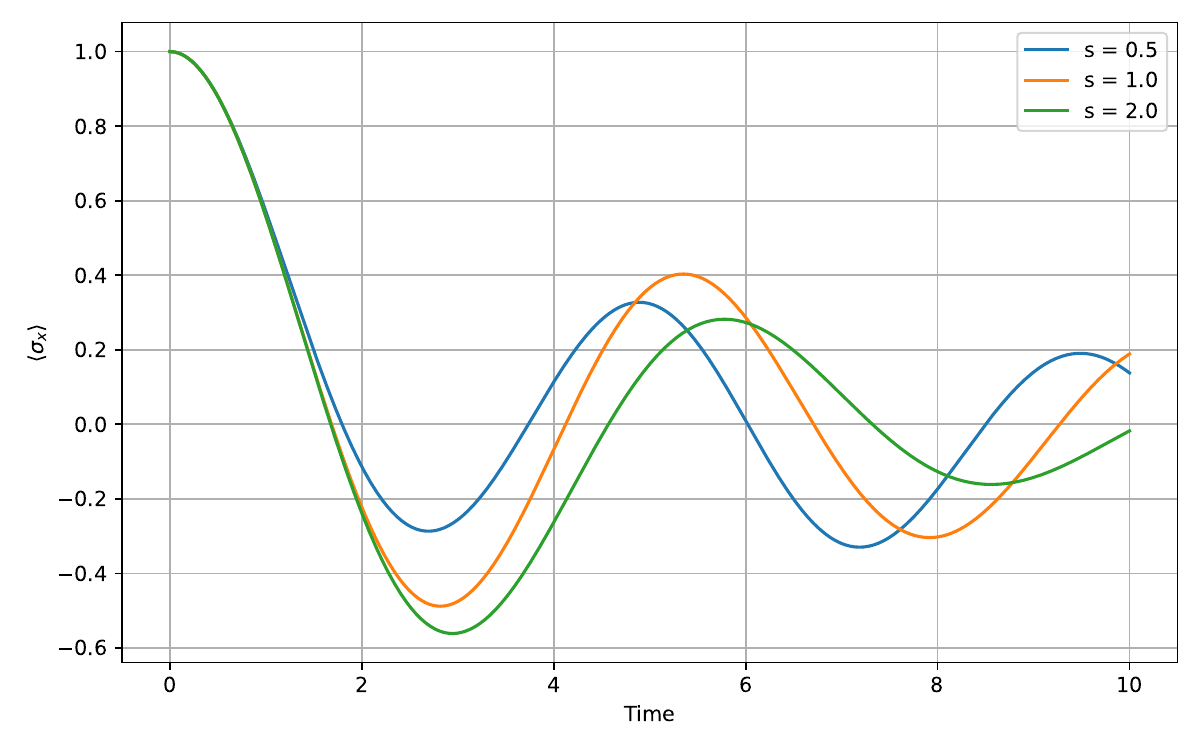}
\caption{Benchmarking the  pure dissipation sector. Temporal evolution of
$\av{\sigma_x}_t$ for a purely dissipative environment, obtained from the
numerical implementation by setting
$J_{xz}(\omega)=J_{zx}(\omega)=J_{zz}(\omega)=0$, such that only
$J_{xx}(\omega)$ contributes. The spectral-density parameters are
$\eta_x=0.05$ and $\omega_{cx}=0.5$, with representative values of $s_x$
in the sub-Ohmic ($s_x<1$), Ohmic ($s_x=1$), and super-Ohmic ($s_x>1$)
regimes. The resulting profiles may be qualitatively compared with those
reported in Ref.~\cite{Barr2024}.}
    \label{fig:2}
\end{figure}

\textbf{Qualitative Benchmark against  pure dissipation Spin-Boson Dynamics:} 
 As a second validation of the numerical implementation, we consider the
amplitude-damping spin-boson dynamics studied in
Ref.~\cite{Barr2024}, where the evolution of $\av{\sigma_x}_t$ was used
to distinguish sub-Ohmic, Ohmic, and super-Ohmic environments. The
corresponding transverse-coupling limit of the present two-channel model
is obtained by setting
\begin{eqnarray}
J_{xx}(\omega)=J(\omega),\qquad
J_{zz}(\omega)=J_{xz}(\omega)=J_{zx}(\omega)=0.
\end{eqnarray}
We characterize the transverse bath by the Ohmic-type spectral density
\begin{eqnarray}
J(\omega)
&=&
\eta_x\,
\omega_{cx}^{1-s_x}\,
\omega^{s_x}
\exp\left(-\frac{\omega}{\omega_{cx}}\right),
\label{eq:jxx}
\end{eqnarray}
where $s_x<1$, $s_x=1$, and $s_x>1$ correspond to sub-Ohmic, Ohmic,
and super-Ohmic spectra, respectively. We compute the dynamics of
$\av{\sigma_x}_t$ for
$\eta_x=0.05$, $\omega_{cx}=0.5$, $\beta=0.1$, and $\omega_0=1$,
considering the representative values
$s_x=0.5$, $1.0$, and $2.0$. The resulting evolution profile,
presented in Fig.\ \ref{fig:2} exhibit
the same qualitative dependence on the Ohmicity parameter as those
reported in Ref.~\cite{Barr2024}. Because the spectral-density
normalization and parameter conventions are not identical, we do not
seek a pointwise numerical reproduction. Rather, this comparison serves
as a qualitative benchmark confirming that the general two-channel
formulation consistently reduces to the standard transverse-coupling
amplitude-damping limit.

\section{Numerical Results: Cross-Spectral Effects on Coherence and Population Dynamics} 
\label{sec:cross_effects}

\begin{figure}[t]
    \centering
\begin{subfigure}[t]{0.48\columnwidth}
\centering
\includegraphics[width=\linewidth]{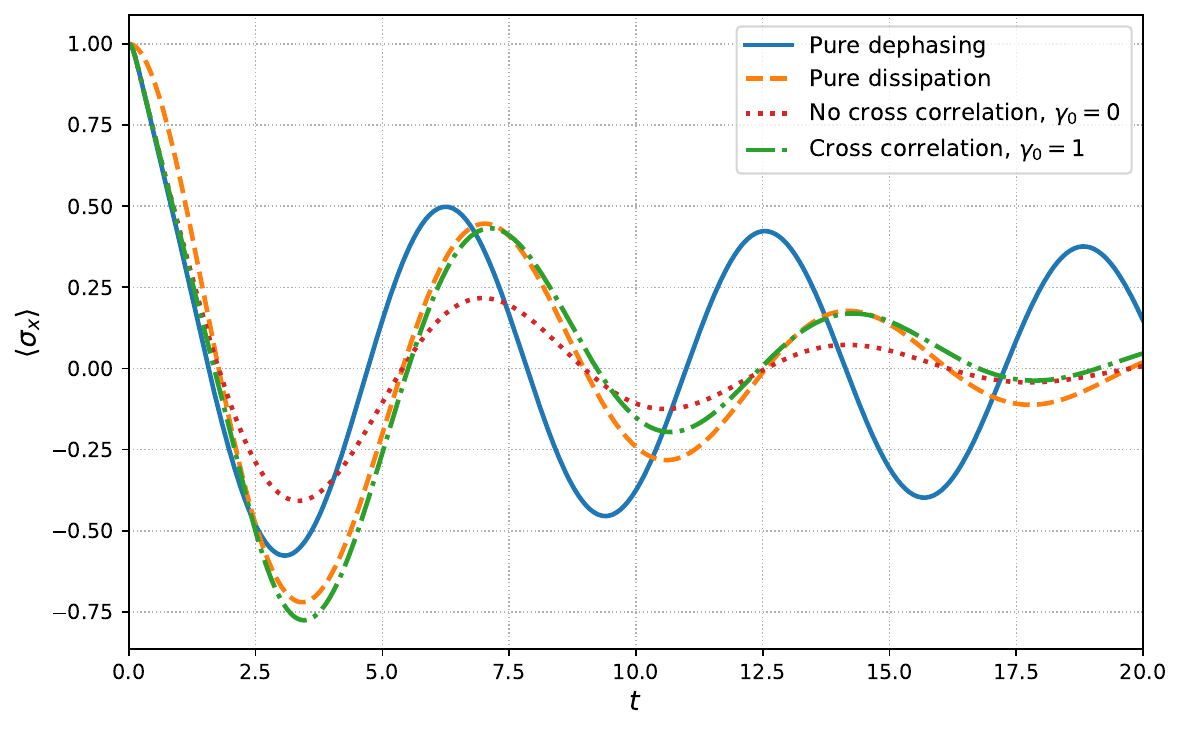}
\caption{Temporal profile of $\av{\sigma_z}_t$}
\label{fig:3a}
\end{subfigure}
	\hfill
\begin{subfigure}[t]{0.48\columnwidth}
\centering
\includegraphics[width=\linewidth]{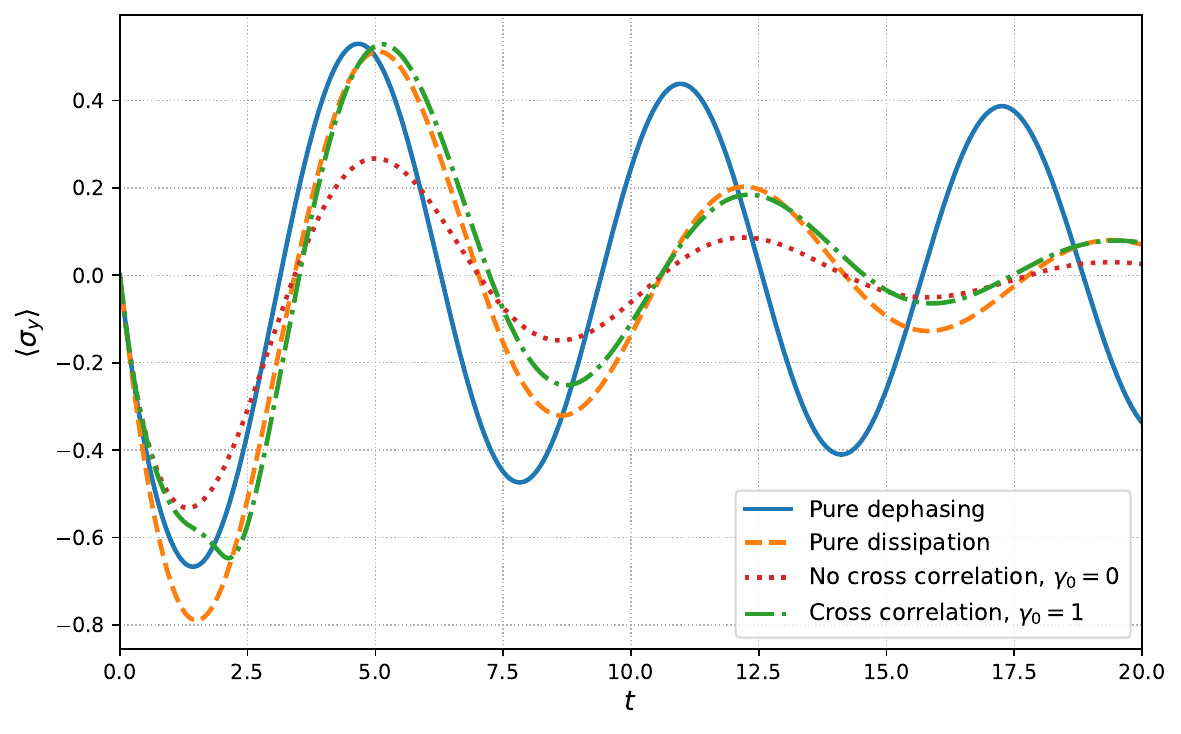}
\caption{Temporal profile of $\av{\sigma_y}_t$}
\label{fig:3b}
\end{subfigure}
    
\begin{subfigure}[t]{0.48\columnwidth}
\centering
\includegraphics[width=\linewidth]{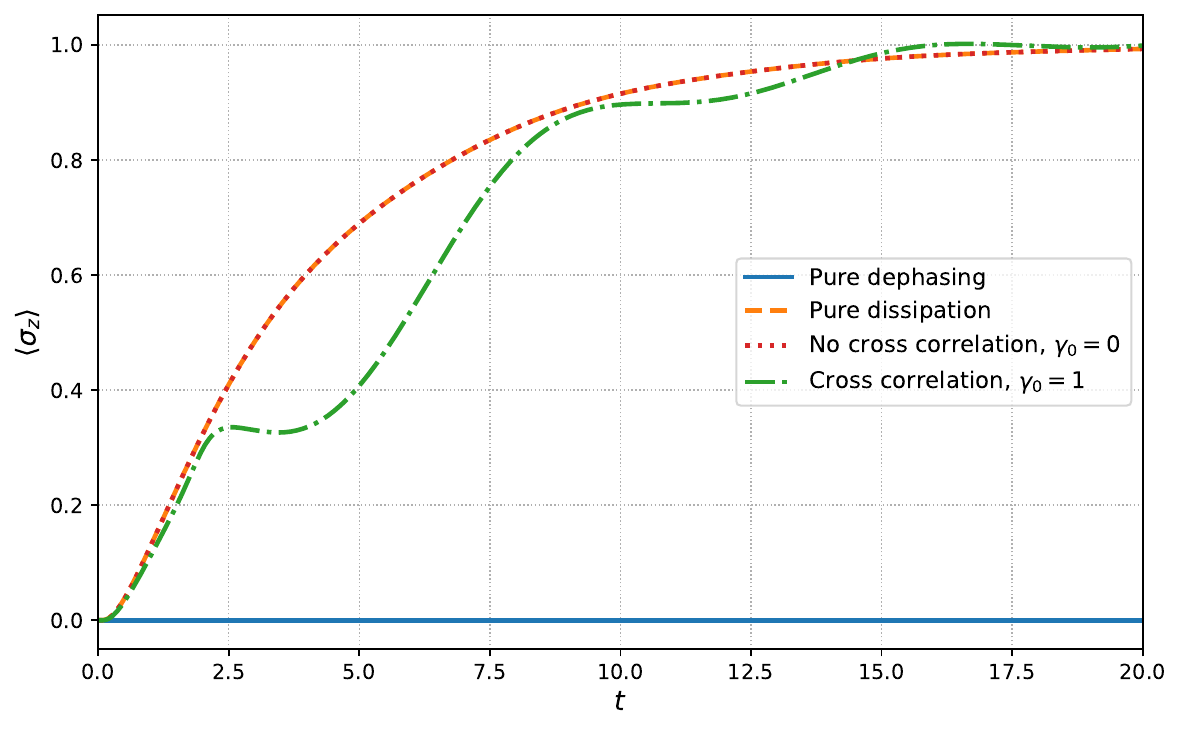}
\caption{Temporal profile of $\av{\sigma_z}_t$}
\label{fig:3c}
\end{subfigure}
    \hfill
\begin{subfigure}[t]{0.48\columnwidth}
\centering
\includegraphics[width=\linewidth]{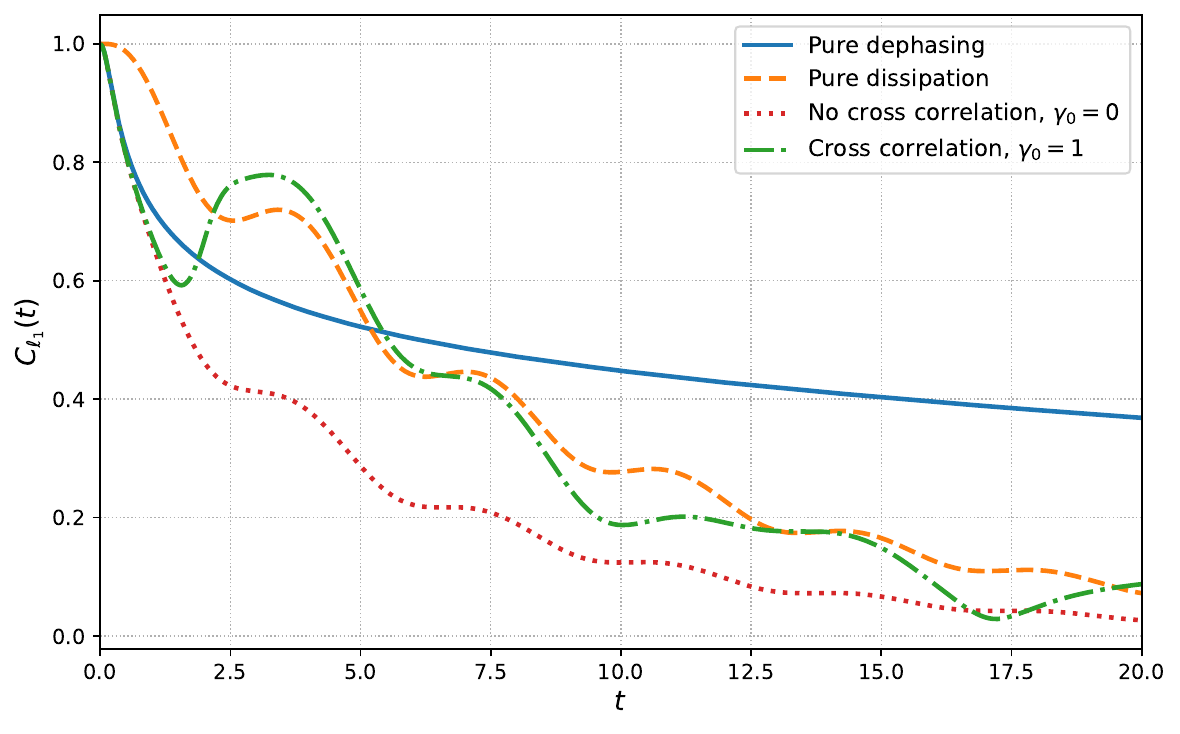}
\caption{Temporal profile of coherence $C(t)$}
\label{fig:3d}
\end{subfigure} 
\caption{Temporal evolution of the Bloch-vector components of $\rho(t)$ and coherence $C(t)$ for the initial state $\rho(0)=\ket{+}\bra{+}$. The parameter sets are - for pure dephasing: $(\eta_z=0.05$, $s_z=1$, $\omega_{cz}=5,\eta_x=0)$; pure dissipation -  $(\eta_x=0.05$, $s_x=1$, $\omega_{cx}=5, \eta_z=0)$; simultaneous dephasing and dissipation without cross correlations, ($\eta_x=\eta_z=0.05$, $s_x=s_z=1$, $\omega_{cx}=\omega_{cz}=5,\gamma_0=0)$; and cross-correlated dephasing and dissipation, with the same parameters but $\gamma_0=1$ and $\tau=2$.
For all plots we take $\beta=40$.}
    \label{fig:3}
\end{figure}

Having validated the longitudinal and transverse sectors independently, we now consider the full two-channel dynamics in the simultaneous presence of dephasing and relaxation environments. In addition to the diagonal auto-spectral
densities $J_{zz}(\omega)$ and $J_{xx}(\omega)$, the environmental
correlations between the two channels are  encoded in the cross-spectral densities $J_{xz}(\omega)$ and $J_{zx}(\omega)$.
 Hermiticity of the bath
correlation matrix requires $J_{zx}(\omega)=J_{xz}^{\ast}(\omega)$.
The completely positive full spectral-density matrix $\mathbf{J}(\omega)$,
defined in  \eqref{eq:Jform}, is parametrized by the auto-spectral densities $J_{xx}(\omega)$ and $J_{zz}(\omega)$, together with the complex correlation coefficient $\gamma(\omega)=\gamma_{0}e^{-\omega/\omega_c}e^{i\theta(\omega)}$ which determines the cross-spectral densities $J_{xz}(\omega)$ and $J_{zx}(\omega)$ through Eq.\ (\ref{eq:Jcross}). The parameter
$0\leq\gamma_{0}\leq 1$  quantifies the strength of the cross correlation
and $\omega_c$ sets the  frequency cut-off  
 beyond which the correlations are exponentially suppressed.
We choose the frequency-dependent phase as $\theta(\omega)=\omega\tau+\phi$, where $\tau$ and $\phi$ respectively characterizes  the relative time delay between 
two correlated noise channels
and the frequency-independent phase offset between them.  \\

The dynamics corresponding to uncorrelated dephasing and relaxation channels
is recovered by setting
$\gamma_0=0$, for which $J_{xz}(\omega)=J_{zx}(\omega)=0$. 
This case provides the reference evolution against which the effects of
cross correlations are assessed. The diagonal
spectral-density parameters are kept fixed when comparing correlated and
uncorrelated environments. Consequently, any difference between the two
evolutions can be attributed directly to the cross-spectral contributions
rather than to a change in the individual dephasing or relaxation spectra.\\

In the absence of cross correlations, the longitudinal and transverse
environmental contributions enter the dynamics only through their respective
auto-correlation functions. When $J_{xz}(\omega)$ and $J_{zx}(\omega)$ are
nonzero, additional terms appear in the TCL2 generator that couple the
dephasing and relaxation sectors. These terms modify both the damping and
the phase evolution of the Bloch-vector components. Consequently, the
correlated dynamics cannot, in general, be reproduced by simply adding the
independent effects of pure dephasing and transverse relaxation.\\

 Figs.~\ref{fig:3a}, \ref{fig:3b}, and \ref{fig:3c} depict the temporal evolution of the Bloch-vector components $\av{\sigma_x}_t$, $\av{\sigma_y}_t$, and $\av{\sigma_z}_t$, respectively, for a representative choice of spectral densities. 
The influence of the cross-spectral densities is illustrated in Fig.~\ref{fig:3} by comparing the dynamics for the maximally correlated case, $\gamma_0=1$, with those for the uncorrelated case, $\gamma_0=0$. In this comparison, the diagonal spectral densities $J_{xx}(\omega)$ and $J_{zz}(\omega)$ are kept fixed and are taken to have the form given in Eq.~\eqref{eq:Jdiag}, with $\eta_x=\eta_z=0.05$, $s_x=s_z=1$, $\omega_{cx}=\omega_{cz}=5$
and frequency-independent phase offsets $\phi=0$.  
The plots for 
 cross-correlated dynamics correspond to 
 cross-correlation cutoff frequency $\omega_c=5$ and relative delay parameter $\tau=2$.   For reference, the corresponding pure-dephasing and pure-relaxation dynamics are also included using the same channel-specific parameter values. The resulting coherence dynamics for 
the different scenarios are shown in Fig.~\ref{fig:3d}. \\

The population dynamics captured in $\av{\sigma_z}_t = \rho_{00}(t) - \rho_{11}(t)$
exhibit a clear distinction between the correlated
and uncorrelated environments. For pure dephasing,
$\av{\sigma_z}_t$ remains equal to its initial value, as expected, since a
longitudinal coupling does not induce transitions between the qubit energy
levels. In contrast, the purely transverse environment drives
$\av{\sigma_z}_t$ toward its stationary positive-$z$ value. The curves for
pure dissipation and for simultaneous dephasing and dissipation without
cross correlations are indistinguishable
implying additional longitudinal dephasing
channel does not   modify the population relaxation generated by
the transverse bath. The inclusion of cross correlations qualitatively changes this behavior.
The evolution of $\av{\sigma_z}_t$ becomes non-monotonic, displaying a
temporary suppression of population relaxation at intermediate times,
followed by a subsequent recovery toward the asymptotic value. Thus, the
cross-spectral terms do not merely change the overall relaxation rate, but
introduce a time-dependent modulation of the population transfer. This
behavior results from the coupling, through $J_{xz}(\omega)$ and
$J_{zx}(\omega)$, of the otherwise distinct longitudinal-dephasing and
transverse-relaxation sectors.\\

The influence of the cross correlations is even more apparent in the
$\ell_1$-norm coherence $
C(t)=\sqrt{\av{\sigma_x}_t^2+\av{\sigma_y}_t^2}$.
When both channels are present but uncorrelated, the coherence decays more
rapidly than in either of the individual-channel cases, reflecting the
combined action of dephasing and relaxation. Although weak oscillatory
features remain, the overall coherence is strongly suppressed.\\

For the cross-correlated environment, the initial coherence decay is
followed by a pronounced transient revival. In particular, after reaching
an early local minimum, $C(t)$ increases over a finite time
interval and attains values substantially larger than those obtained in
the uncorrelated case. Smaller oscillatory enhancements are also visible
at later times. The cross correlations therefore redistribute the
decoherence in time: they can partially counteract the combined damping
produced by the diagonal spectra during some time intervals, while
enhancing the decay during others. Consequently, cross correlations do not
provide a uniform suppression of decoherence, but generate alternating
regimes of coherence preservation and coherence loss.\\

Since $C(t)=\sqrt{\av{\sigma_x}_t^2+\av{\sigma_y}_t^2}$ is the magnitude of the transverse Bloch-vector projection, a coherence revival corresponds to a transient increase in the transverse radius. The cross-spectral terms may also modify the azimuthal orientation of the Bloch vector by redistributing weight between $\av{\sigma_x}_t$ and $\av{\sigma_y}_t$. Examining the two components separately therefore distinguishes a genuine coherence enhancement from a correlation-induced rotation in the $x$--$y$ plane. \\

The small overshoot of $\av{\sigma_z}_t$ beyond its physically allowed
range near the late-time maximum should not be interpreted as a physical
population enhancement. If it exceeds the numerical integration tolerance,
it indicates a loss of positivity of the TCL2 solution and hence the onset
of a regime in which the second-order perturbative approximation is no
longer quantitatively reliable. The corresponding parameter set must
therefore be checked by monitoring the eigenvalues of $\rho(t)$ and the
Bloch-vector norm $\sqrt{\av{\sigma_x}_t^2 + \av{\sigma_y}_t^2 + \av{\sigma_z}_t^2} \leqslant 1$.  \\

The parameter values defining the cross-spectral density
$J_{xz}(\omega)$  determine  the extent to which environmental cross correlations modify the qubit dynamics relative to the case of uncorrelated dephasing and dissipation channels. We investigate such
correlation induced modifications 
of the population imbalance $\av{\sigma_z}_t$ and   coherence $C(t)$ which 
provide  a compact characterization of the two principal effects of the
environment on the qubit.  
While  $C(t)$ quantifies
the preservation or revival of phase coherence,
 $\av{\sigma_z}_t=\rho_{00}(t)-\rho_{11}(t)$  quantifies 
 the redistribution
of population between the qubit levels and directly tracks the mean qubit
energy, thereby revealing correlation-induced modifications of relaxation/
excitation, and energy exchange with the environment. \\

% Although $\av{\sigma_z}_t$ remains constant under pure dephasing and evolves only in the presence of transverse dissipation, its relaxation profile is further modified by the cross-spectral contributions in a correlated dephasing--dissipation environment, as is evident from 
% Fig.~\ref{fig:3c}.  \\
  
  To quantify the cross-correlation-induced modifications, we define
\begin{eqnarray}
\Delta C(t)
&=&
C_{\mathrm{corr}}(t)-C_{\mathrm{uncorr}}(t),
\label{eq:coherence_difference}
\\
\Delta\av{\sigma_z}_t
&=&
\av{\sigma_z}_{t,\mathrm{corr}}
-
\av{\sigma_z}_{t,\mathrm{uncorr}}.
\label{eq:imbalance_difference}
\end{eqnarray}
The subscripts ``corr'' and ``uncorr'' denote dynamics with and
without cross-spectral densities, respectively. A positive
$\Delta C(t)$ signifies correlation-induced coherence enhancement.  
A positive value of $\Delta\av{\sigma_z}_t$
 indicates that the relaxation of qubit  toward its ground state
 is more enhanced than in the uncorrelated case and therefore retains less energy. By contrast, a negative value signifies suppressed relaxation, a larger residual excited-state population, and consequently greater energy retention in the qubit.  \\ 
 
\begin{figure}[t]
    \centering
    \begin{subfigure}[t]{0.48\columnwidth}
        \centering
        \includegraphics[width=\linewidth]{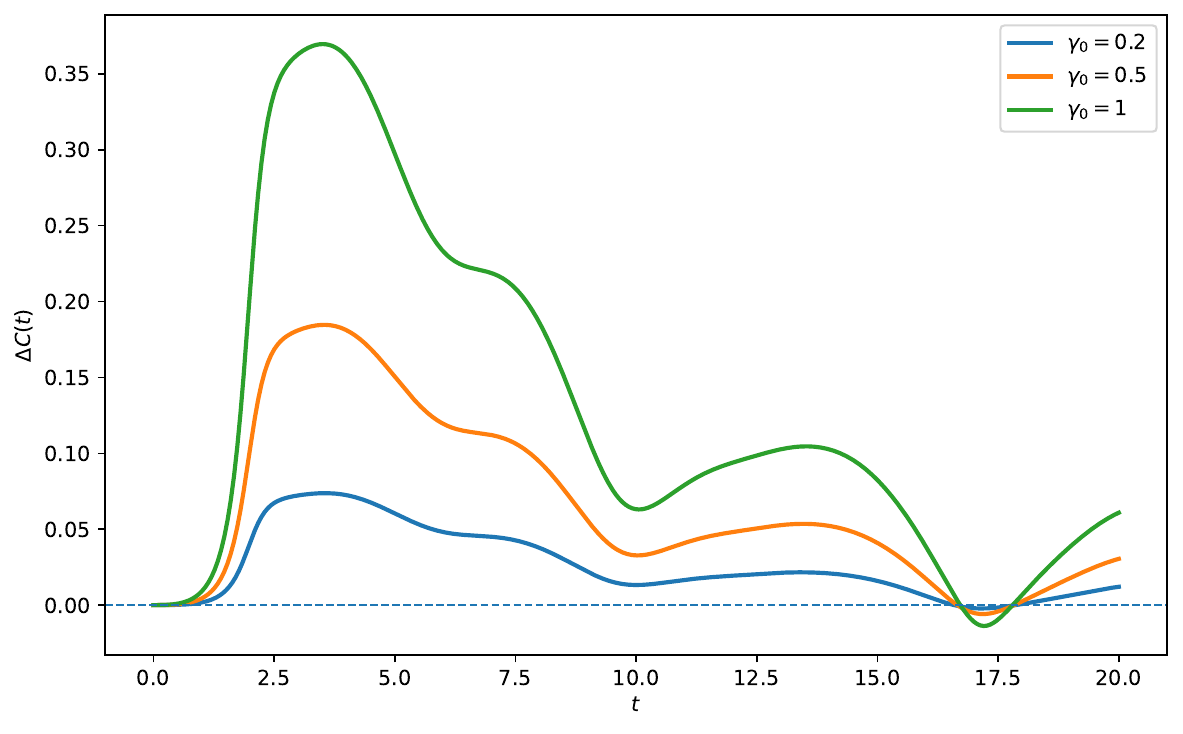}
    \end{subfigure}
    \hfill
    \begin{subfigure}[t]{0.48\columnwidth}
        \centering
        \includegraphics[width=\linewidth]{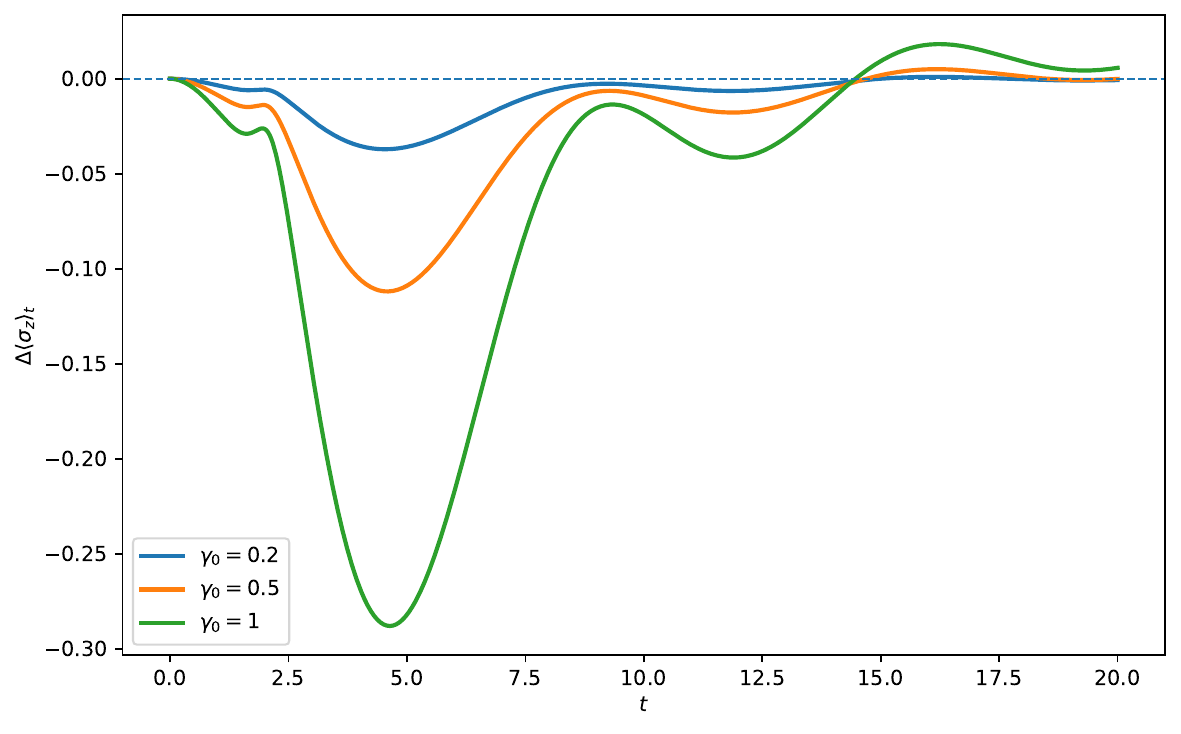}
    \end{subfigure}
\caption{Temporal evolution of $\Delta C(t)$ (left panel) and
$\Delta\av{\sigma_z}_t$ (right panel) for representative cross-correlation strengths $\gamma_0=0.2$, $0.5$, and $1.0$. The remaining cross-spectral parameters are fixed at $\omega_c=5,\tau=2$ and $\phi=0$, with inverse temperature $\beta=40$. The diagonal spectral-density parameters are $\eta_x=\eta_z=0.05$, $s_x=s_z=1$, and $\omega_{cx}=\omega_{cz}=5$.}
    \label{fig:4}
\end{figure}

\begin{figure}[t]
    \centering
    \begin{subfigure}[t]{0.48\columnwidth}
        \centering
        \includegraphics[width=\linewidth]{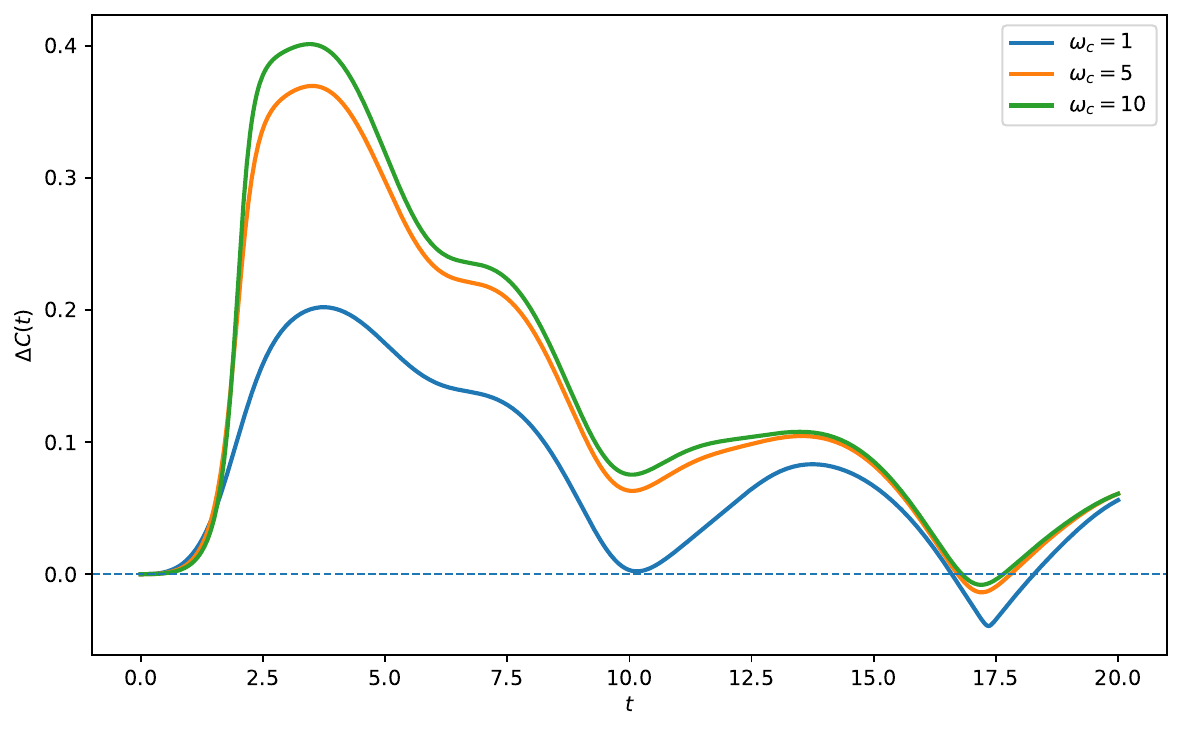}
    \end{subfigure}
    \hfill
    \begin{subfigure}[t]{0.48\columnwidth}
        \centering
        \includegraphics[width=\linewidth]{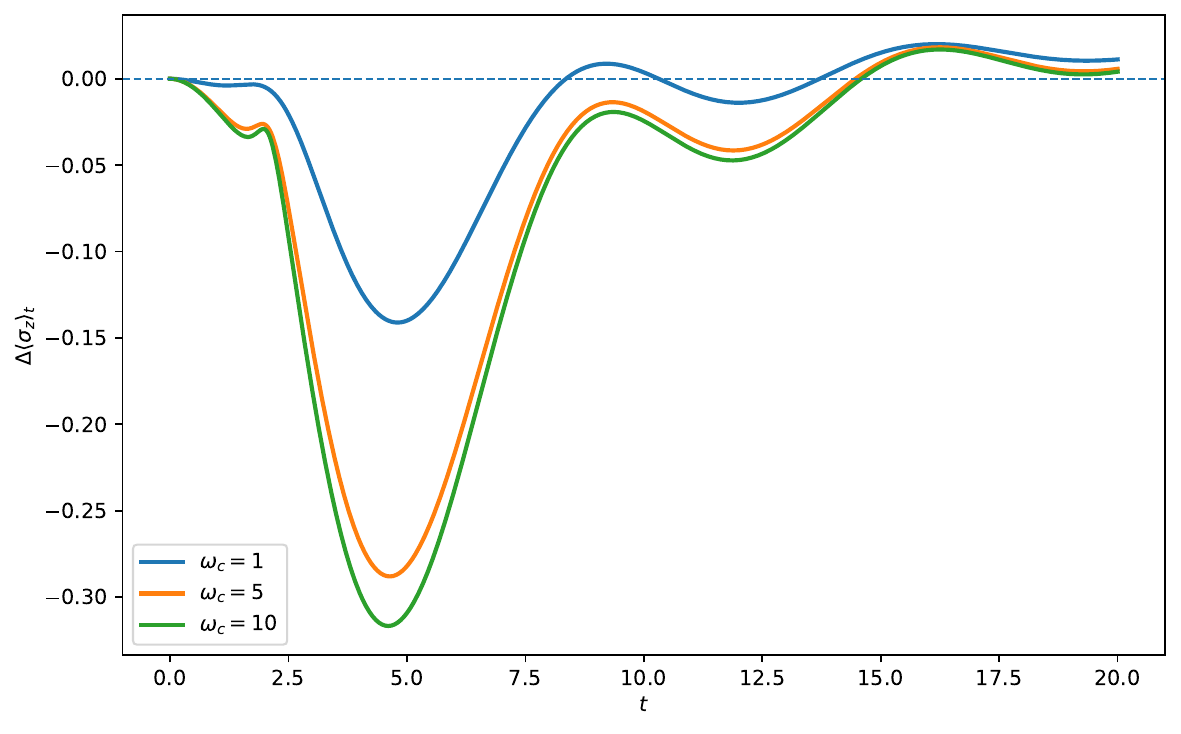}
    \end{subfigure}
\caption{Temporal evolution of $\Delta C(t)$ (left panel) and
$\Delta\av{\sigma_z}_t$ (right panel) for representative 
values of the cross-spectral cutoff frequency, 
$\omega_{c}=1$, $5$, and $10$. The
remaining cross-spectral parameters are fixed at $\gamma_0=1, \tau=2$ and $\phi=0$, with inverse temperature $\beta=40$. The diagonal
spectral-density parameters are 
$\eta_x=\eta_z=0.05$, $s_x=s_z=1$, and
$\omega_{cx}=\omega_{cz}=5$.}
\label{fig:5}
\end{figure}

\begin{figure}[t]
    \centering
    \begin{subfigure}[t]{0.48\columnwidth}
        \centering
        \includegraphics[width=\linewidth]{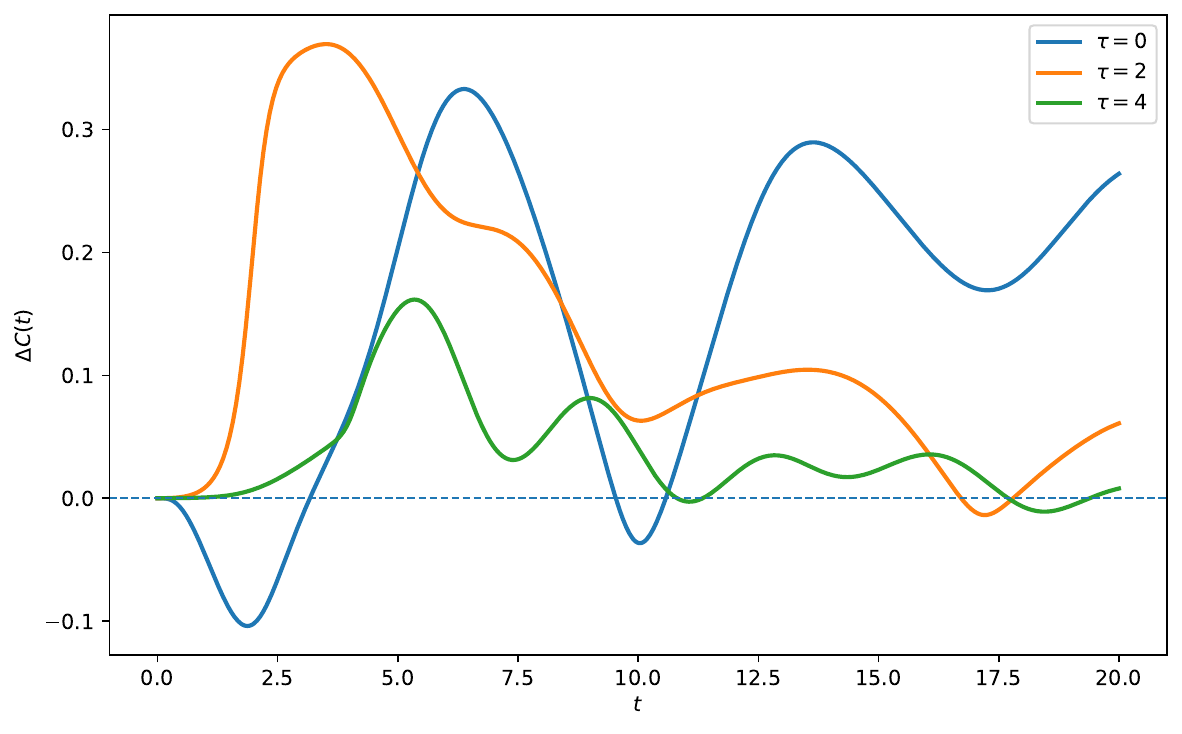}
    \end{subfigure}
    \hfill
    \begin{subfigure}[t]{0.48\columnwidth}
        \centering
        \includegraphics[width=\linewidth]{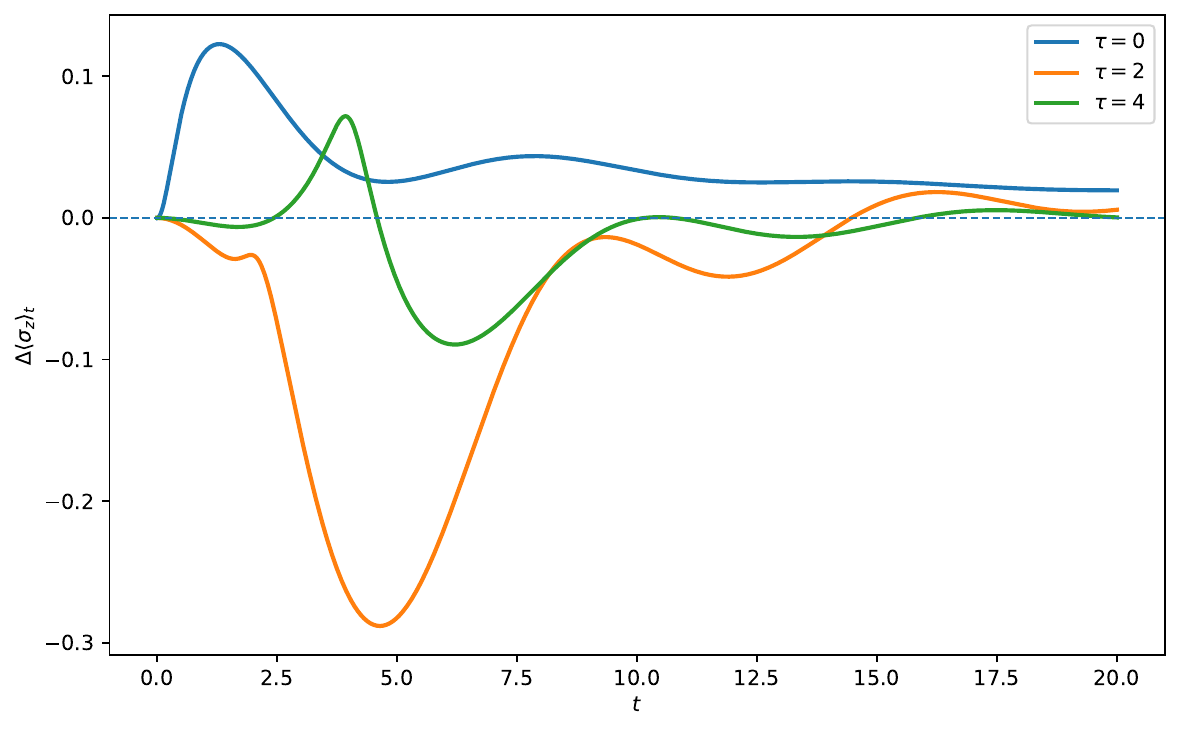}
    \end{subfigure}
\caption{Temporal evolution of $\Delta C(t)$ (left panel) and
$\Delta\av{\sigma_z}_t$ (right panel) for representative values 
relative time delay, $\tau=0$, $2$, and $4$ between the two correlated noise channels. The remaining
cross-spectral parameters are fixed at $\gamma_0=1, 
\omega_c=5$ and $\phi=0$, with inverse temperature $\beta=40$. 
The diagonal
spectral-density parameters are 
$\eta_x=\eta_z=0.05$, $s_x=s_z=1$, and
$\omega_{cx}=\omega_{cz}=5$.}
\label{fig:6}
\end{figure}

\begin{figure}[t]
    \centering
    \begin{subfigure}[t]{0.48\columnwidth}
        \centering
        \includegraphics[width=\linewidth]{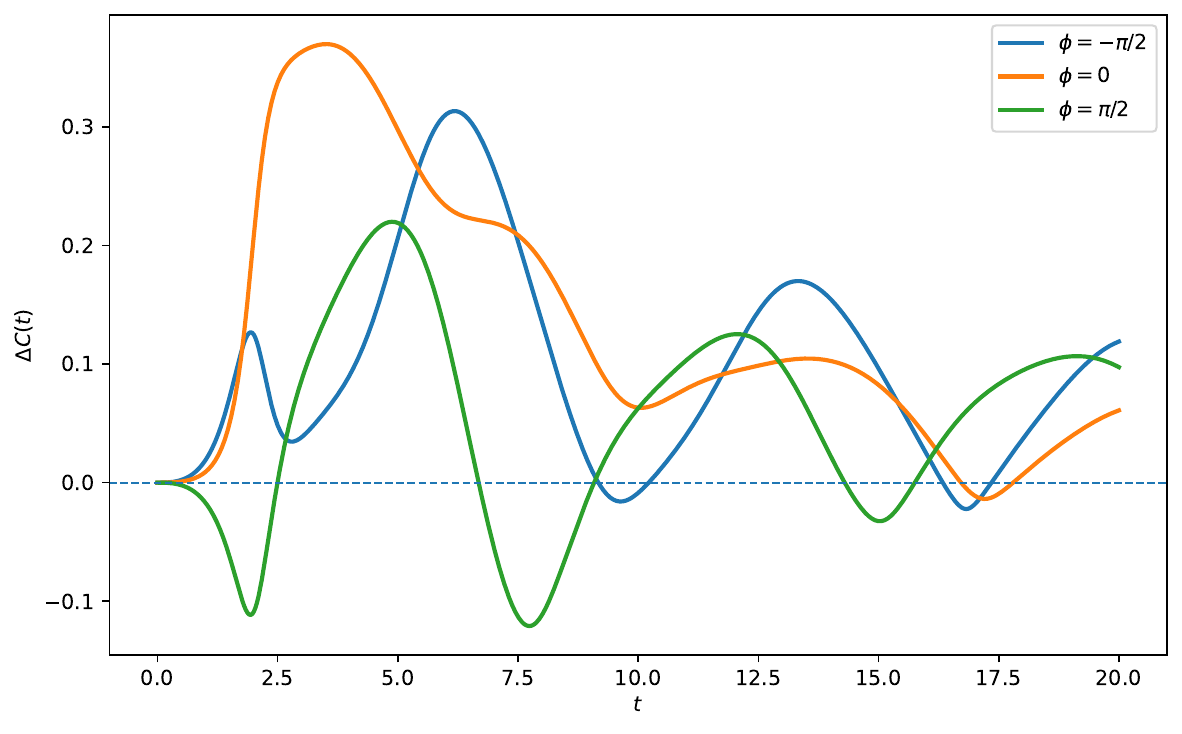}
    \end{subfigure}
    \hfill
    \begin{subfigure}[t]{0.48\columnwidth}
        \centering
        \includegraphics[width=\linewidth]{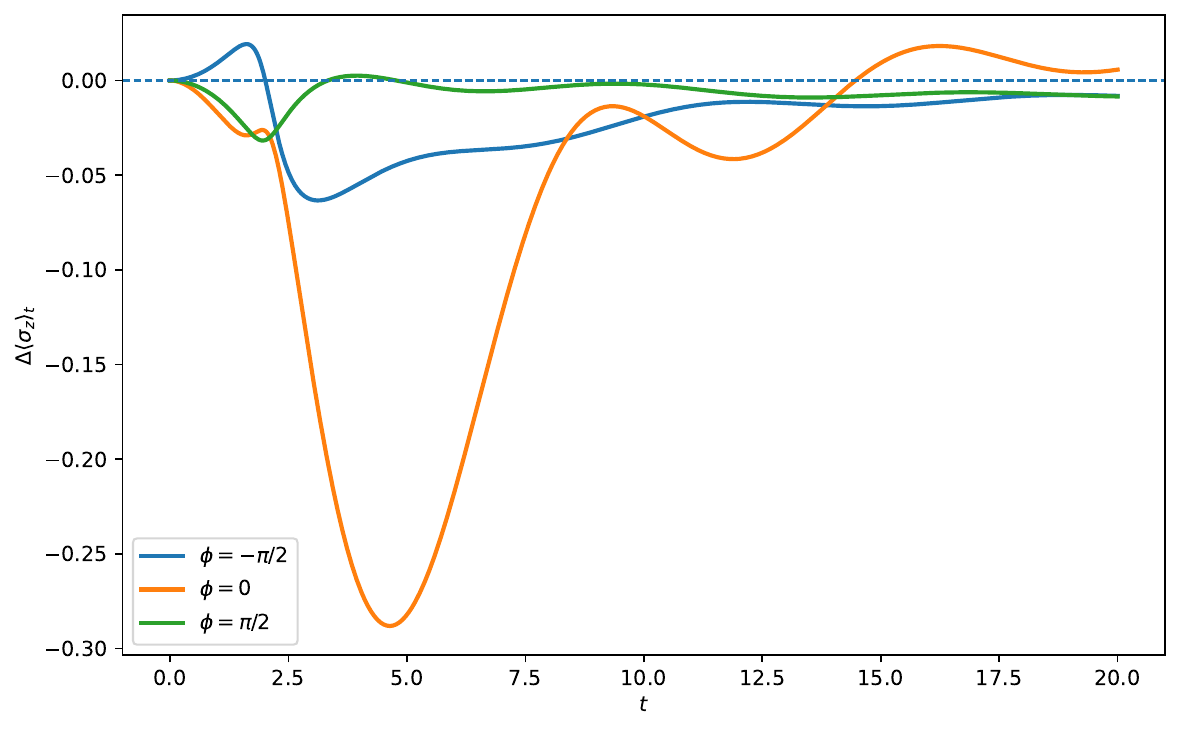}
    \end{subfigure}
\caption{Temporal evolution of $\Delta C(t)$ (left panel) and
$\Delta\av{\sigma_z}_t$ (right panel) for  different frequency-independent phase offsets $\phi=-\pi/2,0,\pi/2$ between the two correlated noise channels. The remaining
cross-spectral parameters are fixed at $\gamma_0=1, \tau=2$ and
$\omega_c=5$,  with inverse temperature $\beta=40$. 
The diagonal
spectral-density parameters are 
$\eta_x=\eta_z=0.05$, $s_x=s_z=1$, and
$\omega_{cx}=\omega_{cz}=5$.}
\label{fig:7}
\end{figure}

\begin{figure}[t]
    \centering
    \begin{subfigure}[t]{0.48\columnwidth}
        \centering
        \includegraphics[width=\linewidth]{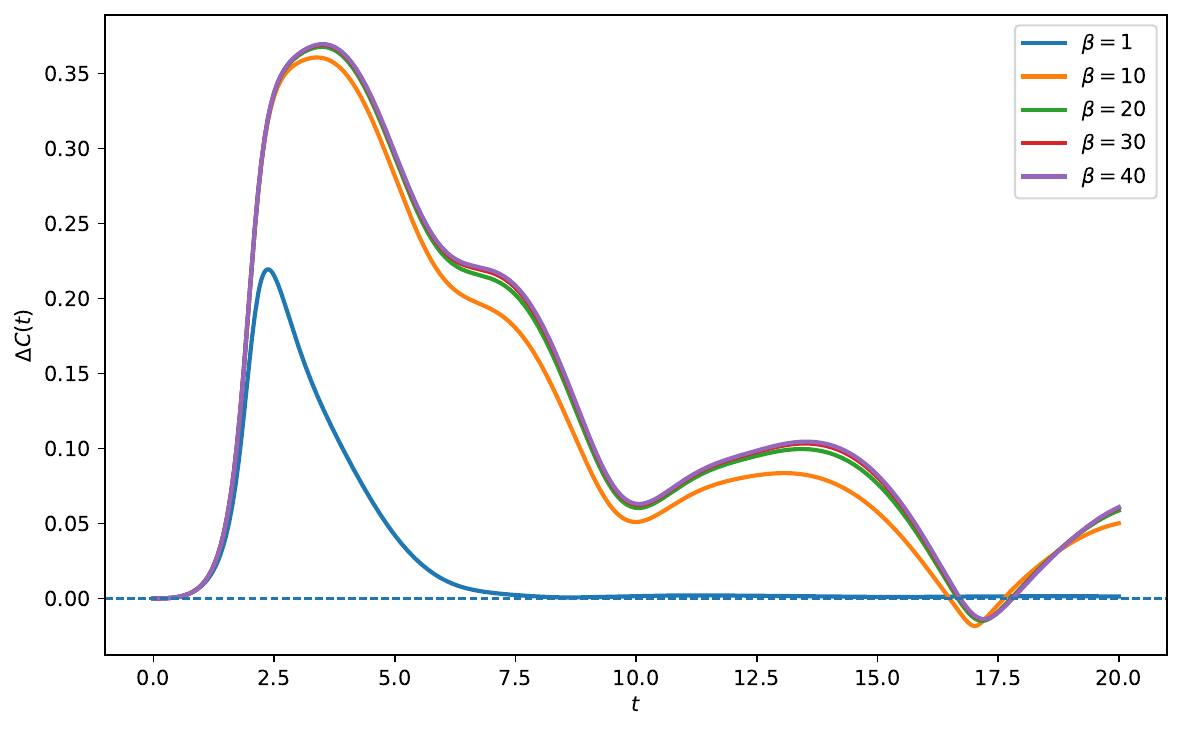}
    \end{subfigure}
    \hfill
    \begin{subfigure}[t]{0.48\columnwidth}
        \centering
        \includegraphics[width=\linewidth]{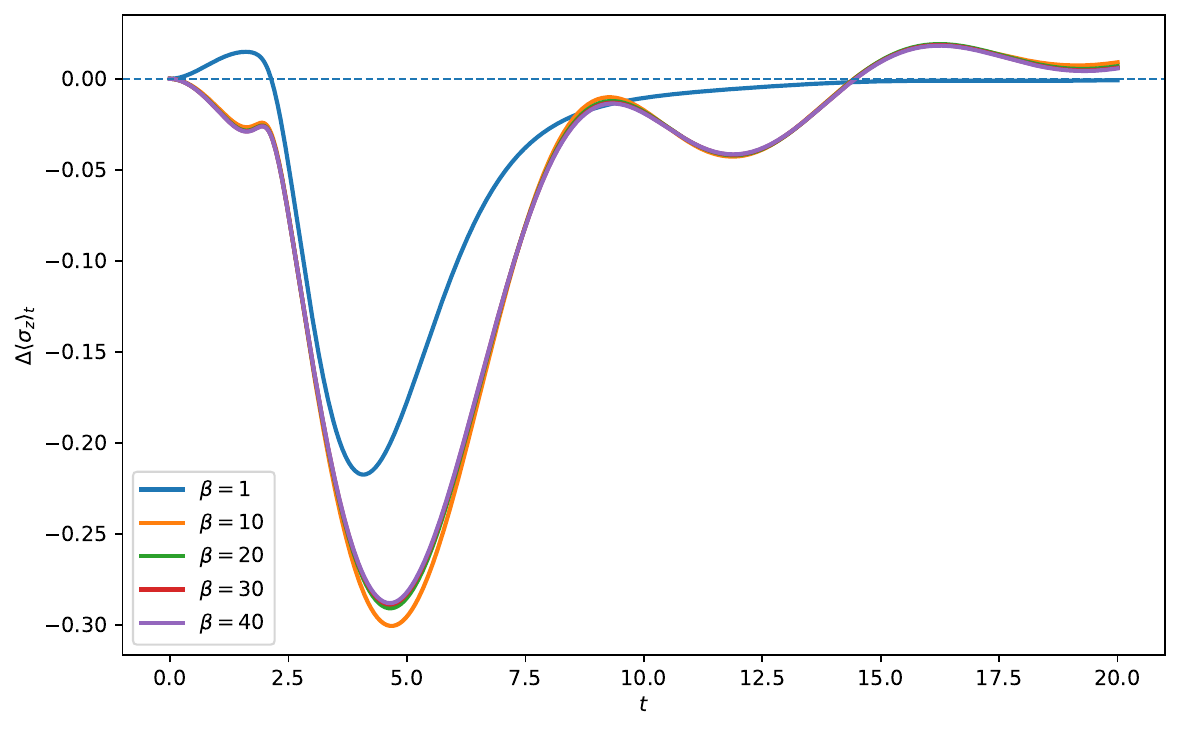}
    \end{subfigure}
\caption{Temporal evolution of $\Delta C(t)$ (left panel) and
$\Delta\av{\sigma_z}_t$ (right panel) for  different inverse temperatures. The  
cross-spectral parameters are fixed at $\gamma_0=1, \tau=2$ and
$\omega_c=5$, $\phi=0$.
The diagonal
spectral-density parameters are 
$\eta_x=\eta_z=0.05$, $s_x=s_z=1$, and
$\omega_{cx}=\omega_{cz}=5$.}
\label{fig:8}
\end{figure}

For numerical analysis of cross-spectral effects we keep the diagonal auto-spectral densities
$J_{xx}(\omega)$ and $J_{zz}(\omega)$ identical and if
of Ohmic type fixing $\eta_x=\eta_z=0.05$, $s_x=s_z=1$, and
$\omega_{cx}=\omega_{cz}=5$. Unless explicitly varied, the cross-spectral
parameters are fixed at $\gamma_0=1$, $\tau=2$, $\phi=0$, and
$\omega_c=\ldots$, while the inverse temperature is set to $\beta=40$.
 We examine the dependence of the correlation-induced dynamics of
 $\Delta C(t)$ and $\av{\sigma_z}_t$ on the
cross-correlation strength $\gamma_0$ in
Fig.~\ref{fig:4}; the cross-spectral cutoff frequency
$\omega_c$  in Fig.~\ref{fig:5}; the relative delay
$\tau$  in Fig.~\ref{fig:6}; the phase offset
$\phi$ in Fig.~\ref{fig:7}; and the inverse
temperature $\beta$ in Fig.~\ref{fig:8}. \\

 As shown in Fig.~\ref{fig:4}, increasing the cross-correlation strength $\gamma_0$ systematically amplifies the correlation-induced changes in the dynamics. In particular, the dominant positive peak of $\Delta C(t)$ becomes more pronounced, indicating a stronger enhancement of coherence with increasing $\gamma_0$. Simultaneously, the negative minimum of $\Delta\av{\sigma_z}_t$ deepens at intermediate times, showing that the cross correlations increasingly slow the relaxation toward the ground state. The qubit therefore retains a larger excited-state population and, correspondingly, a higher average energy than in the uncorrelated case.  \\
 
Fig.~\ref{fig:5} shows that increasing $\omega_c$ enhances both the correlation-induced coherence gain, as reflected in $\Delta C(t)$, and the suppression of relaxation relative to the uncorrelated case, as reflected in $\Delta\av{\sigma_z}_t$. This behavior arises because a larger $\omega_c$ extends the frequency interval over which the longitudinal and transverse environmental channels remain appreciably correlated.
The relatively small difference between the curves for $\omega_c=5$ and $\omega_c=10$ indicates that these effects are approaching saturation, as the cross-spectral bandwidth already covers the frequency region where the diagonal spectral densities are most significant.\\ 

The results shown in Figs.~\ref{fig:6} and \ref{fig:7} reveal a pronounced sensitivity of the cross-correlated dynamics to the phase parameters. 
Changing either the delay $\tau$ or the frequency-independent phase offset 
$\phi$ shifts the positions of the extrema and may reverse the signs of both 
$\Delta C(t)$ and $\Delta\av{\sigma_z}_t$. Consequently, the cross-spectral 
contribution can either enhance or suppress coherence and relaxation, 
depending on whether the correlated environmental fluctuations interfere 
constructively or destructively. For the benchmark parameters considered 
here, $\phi=0$ yields the strongest early-time enhancement of coherence 
and the most pronounced suppression of relaxation relative to the uncorrelated case, 
whereas $\phi=\pm\pi/2$ weakens or delays these effects 
and may lead to an oscillatory response.\\

In addition to the parameters specifying the spectral densities, the bath temperature, ($\beta^{-1}$), provides an independent means of controlling the dynamics through the thermal weighting of the bath modes entering the matrix elements of $\mathbf{M}$ [Eq.~\eqref{eq:Mfinal}], which in turn governs the evolution of the reduced density matrix via Eq.~\eqref{eq:rd15}.  
Figure~\ref{fig:8} shows that the effect  of cross correlations become increasingly pronounced with increasing $\beta$ in the low $\beta$ (high-temperature)
 regime and gradually saturate beyond $\beta\gtrsim10$. 
 %At low $\beta$, both the coherence revival and the associated population modification are strongly suppressed.
  The low-temperature regime (high $\beta$) exhibits a larger and more persistent positive $\Delta C(t)$, accompanied by a more pronounced negative $\Delta\langle\sigma_z\rangle_t$. \\

These results establish that
the magnitude, timing, and even the sign of the correlation-induced
modifications can be controlled through the strength, bandwidth, phase
structure, and temperature dependence of the cross-spectral density.

\section{Operational Significance and Potential Technological Relevance}
\label{sec:tr}
The preceding results demonstrate that cross correlations 
between the dephasing and relaxation channels 
can reshape both coherence decay and
population relaxation. 
Since coherence is a resource for phase-sensitive
quantum operations, whereas relaxation 
limits qubit information-storage
times \cite{Baumgratz2014,Ithier2005}, 
the strength, bandwidth, and phase of
the cross spectrum provide additional 
parameters for controlling the
open-system dynamics beyond simply 
reducing the individual noise strengths.\\

The coherence enhancement admits a direct metrological interpretation in
single-qubit phase estimation. For an unknown phase $\varphi$ encoded by
the unitary rotation generated by $\sigma_z/2$,
$\rho_{\varphi}(t)=e^{-i\varphi\sigma_z/2}\rho(t)e^{i\varphi\sigma_z/2}$,
the quantum Fisher information associated with the estimation of $\varphi$ is \cite{Braunstein1994}
\begin{eqnarray}
F_Q^{(z)}(t)=\av{\sigma_x}_t^2+\av{\sigma_y}_t^2=C^2(t)
\label{eq:qfi_coherence_relation}
\end{eqnarray}
Consequently, $\Delta C(t)>0$ implies
$F_{Q,\mathrm{corr}}^{(z)}(t)>F_{Q,\mathrm{uncorr}}^{(z)}(t)$ and hence,
for a fixed number $\nu$ of independent measurements, a lower quantum
Cram\'er--Rao bound,
$\delta\varphi\geq[\nu F_Q^{(z)}(t)]^{-1/2}$, than in the corresponding
uncorrelated environment \cite{Braunstein1994,Paris2009}. 
Cross-correlation-induced coherence revivals therefore may identify 
finite temporal
windows in which the optimal 
phase-estimation sensitivity exceeds that
attainable in the corresponding uncorrelated environment.\\

The population dynamics provide a complementary operational perspective.
The quantity $\Delta\av{\sigma_z}_t$ measures the cross-correlation-induced change in the population imbalance and, consequently, in the mean qubit energy relative to the uncorrelated case. Its sign indicates whether the correlations enhance relaxation and reduce energy retention or suppress relaxation and preserve a larger excited-state population. 
The negative values of $\Delta\av{\sigma_z}_t$ observed over intermediate
time intervals  indicates suppressed relaxation and a larger
residual excited-state population relative to the uncorrelated case, which
may reduce population-loss errors during storage or sensing intervals
\cite{Ithier2005}.
This excess energy should not, however, be interpreted
directly as a quantum-battery advantage: assessing battery performance
requires an additional analysis of
passivity,
ergotropy, and the energetic cost of 
generating the correlated environment \cite{AlickiFannes2013,Campaioli2017,Ferraro2018}. \\

The parameter dependence shown in Figs.~\ref{fig:4}--\ref{fig:8}
demonstrates that the correlation-induced dynamical windows can be tuned
through the cross-spectral properties. The parameter $\gamma_0$ controls
the overall magnitude of the response, whereas $\omega_c$ determines the
frequency range over which the two environmental channels remain
correlated. The delay $\tau$ and phase offset $\phi$ govern the relative
spectral phase and can shift, enhance, suppress, or reverse the
correlation-induced modification. Temperature provides an additional
control through the thermal weighting of the bath modes. Thus, engineering
the cross spectrum may allow the coherence-enhancement window to be
synchronized with a prescribed sensing, gate, or readout operation, in the
spirit of quantum-reservoir engineering
\cite{Poyatos1996,Murch2012}.\\

The proposed control of the environmental spectra is closely related to reservoir engineering, in which the system--environment coupling is deliberately tailored to generate a desired reduced dynamics \cite{Poyatos1996}. Engineered reservoirs have been experimentally employed to prepare and stabilize coherent and entangled states in superconducting circuits and trapped-ion platforms \cite{Murch2012,Lin2013,Shankar2013}. In the present model, such an implementation would require the two system operators to couple to a common set of spectrally filtered environmental modes. By controlling the relative amplitudes and phases of these couplings, one could, in principle, realize a prescribed cross-spectral density $J_{xz}(\omega)$. The phenomenological parameters $\gamma_0, \omega_c, \tau$ and $\phi$ may therefore be interpreted as target characteristics associated, respectively, with the correlation strength, spectral bandwidth, propagation delay, and relative phase. These features could be engineered through spectral filtering, controlled propagation paths, externally driven auxiliary modes, or coupling to a common resonator \cite{Kronwald2013,Kockum2014,Metelmann2015}.  
The present analysis, however, establishes only a state-level dynamical advantage and does not by itself constitute a complete operational protocol. \\

\section{Conclusion}
\label{sec:conclusion}
We have investigated the reduced dynamics of a qubit subjected to correlated longitudinal and transverse fluctuations arising from a common bosonic environment. The environmental properties were described by a matrix-valued spectral density whose diagonal elements characterize the individual dephasing and relaxation channels, while its complex off-diagonal elements encode correlations between them. Within the second-order time-convolutionless framework, we derived a closed set of time-local equations for the Bloch-vector components, thereby providing a direct description of the coherence, population imbalance, and energy exchange of the qubit.\\

The numerical implementation was validated against the exact 
solution of the pure-dephasing spin-boson model and the 
established qualitative behavior of the transverse-coupling 
amplitude-damping limit. In the simultaneous presence of 
dephasing and relaxation, the cross-spectral terms couple 
the otherwise distinct longitudinal and transverse sectors 
and produce dynamics that cannot be reproduced by simply 
adding the effects of two independent noise channels. In 
particular, the correlations can generate a transient 
revival of coherence following its initial decay and can 
induce a non-monotonic modulation of the population 
relaxation. The magnitude and temporal structure of these 
effects are controlled by the strength and bandwidth of the 
cross correlations, whereas the relative delay and phase 
determine whether the correlated contributions combine 
constructively or destructively. The bath temperature 
provides an additional control parameter through the thermal 
weighting of the environmental modes.\\

These results demonstrate that correlations between
different environmental noise channels may serve as a 
resource for redistributing decoherence and energy 
relaxation over time. They also suggest that suitably 
engineered cross-spectral densities could provide finite 
temporal windows of enhanced coherence or suppressed 
relaxation. The present conclusions remain restricted to the 
weak-coupling regime in which the TCL2 approximation 
preserves the physicality of the reduced state. Future work 
may examine the persistence of these effects using 
nonperturbative methods, characterize the associated 
information-flow properties through independent non-
Markovianity measures, and assess their usefulness for 
quantum sensing, state preservation, and reservoir-
engineering protocols.

\section*{Acknowledgment}
S.D. acknowledges financial support from the University Grants Commission (UGC), Government of
India, through a Senior Research Fellowship (SRF).
S.M. acknowledges financial support from the Council of Scientific and Industrial Research
(CSIR), Government of India, through a Senior Research Fellowship (SRF). 

\bibliographystyle{JHEP}
\bibliography{ref_mod}

\end{document}